\documentclass[final]{article}

\usepackage{graphicx}
\usepackage{comment}
\usepackage{simplemargins}
\usepackage{zed-csp}
\usepackage{IssamCh3}
\usepackage{mythesis}
\usepackage{datarefine}
\usepackage{setspace}
\begin{document}

\renewcommand{\baselinestretch}{1}

\title{The link to the formal publication is via\\ {\small \url{https://doi.org/10.1007/s10766-007-0046-1}}\\ Parallel Algorithms Development for Programmable Devices with Application from Cryptography}

\author{Issam~W.~Damaj,%~\IEEEmembership{Member,~IEEE,}% <-this % stops a space
\thanks{I. Damaj is with Dhofar University, Salalah, Oman, Email:i\_damaj@du.edu.om}}% <-this % stops a space

\maketitle

\begin{abstract}
Reconfigurable devices, such as Field Programmable Gate Arrays
(\textit{FPGAs}), have been witnessing a considerable increase in
density. State-of-the-art \textit{FPGAs} are complex hybrid
devices that contain up to several millions of gates. Recently,
research effort has been going into higher-level parallelization
and hardware synthesis methodologies that can exploit such a
programmable technology. In this paper, we explore the
effectiveness of one such formal methodology in the design of
parallel versions of the \textit{Serpent} cryptographic algorithm.
The suggested methodology adopts a functional programming notation
for specifying algorithms and for reasoning about them. The
specifications are realized through the use of a combination of
function decomposition strategies, data refinement techniques, and
off-the-shelf refinements based upon higher-order functions. The
refinements are inspired by the operators of Communicating
Sequential Processes (\textit{CSP}) and map easily to programs in
\textit{Handel-C} (a hardware description language). In the
presented research, we obtain several parallel \textit{Serpent}
implementations with different performance characteristics. The
developed designs are tested under \textit{Celoxica's}
\textit{RC-1000} reconfigurable computer with its 2 million gates
\textit{Virtex-E} \textit{FPGA}. Performance analysis and
evaluation of these implementations are included.
\end{abstract}

%\begin{keywords}

\section{Introduction}

The rapid progress and advancement in integrated
circuits (\textit{ICs}) technology provides a variety of new
implementation options for system engineers. The choice varies
between the flexible programs running on a general purpose
processor \textit{(GPP)} and the fixed hardware implementation
using an application specific integrated circuit (\textit{ASIC}).
Many other implementation options present, for instance, a system
with a \textit{RISC} processor and a \textit{DSP} core. Other
options include graphics processors and microcontrollers.
Specialized processors certainly improve performance over
general-purpose ones, but this comes as a \textit{quid pro quo}
for flexibility. Combining the flexibility of \textit{GPPs} and
the high performance of \textit{ASICs} leads to the introduction
of reconfigurable computing (\textit{RC}) as a new implementation
option with a balance between versatility and speed.

Field Programmable Gate Arrays (\textit{FPGAs}), are nowadays
important components of \textit{RC}-systems. \textit{FPGAs} have
shown a dramatic increase in their density over the last few
years. For example, companies such as \textit{Xilinx} \cite{126}
and \textit{Altera} \cite{127} have enabled the production of
\textit{FPGAs} with several millions of gates, such as in
\textit{Virtex-II Pro} and \textit{Stratix-II} \textit{FPGAs}. The
versatility of \textit{FPGAs}, opened up completely new avenues in
high-performance computing. These programmable hardware circuits
can be supported with flexible parallel algorithms design
methodologies to form a powerful paradigm for computing.

The traditional implementation of a function on an \textit{FPGA}
is done using logic synthesis based on \textit{VHDL}, Verilog or a
similar \textit{HDL} (hardware description langauge). These
discrete event simulation languages are rather different from
languages, such as \textit{C}, \textit{C++} or \textit{JAVA}. An
interesting step towards more success in hardware compilation was
to grant a high-level of abstraction from the point of view of
programmer. Accordingly, and recently, vendors have initiated the
use of high-level languages like \textit{Handel-C} \cite{129,125},
\textit{Forge} \cite{130}, \textit{Nimble} \cite{131,132}, and
\textit{SystemC} \cite{133}.

Although modern hardware compilation tools have significantly
reduced the complexity of hardware design, many research
opportunities are still present to study even more reduced design
complexity. Accordingly, in this paper we investigate a
methodology enabling high-level of abstraction in the process of
hardware design. The proposed methodology is a step-wise
refinement approach for developing parallel algorithms. The
development will is based on higher-order skeletons exploiting
possible inherent algorithmic parallelism. Algorithmic skeletons
provide a promising basis for the automatic utilization of
parallelism at sites of higher-order functions
\cite{michaelson00nested}. The correctness of the developed
hardware is put forward for further discussion through in this
paper.

\begin{figure}
	\begin{center}
		\includegraphics [scale = 0.8]%[height=2in,width=2in,angle=0]
		{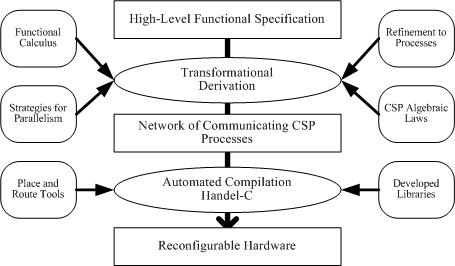}
		\caption{An overview of the transformational derivation and the hardware realisation processes.}
		\label{RDM}
	\end{center}
\end{figure}

The research presented in this paper, builds on the work of
Abdallah and Hawkins \cite{6,9,140,140b} that adopts the
transformational programming approach for deriving massively
parallel algorithms from functional specifications (See
Figure~\ref{RDM}). In this approach, the functional notation is
used for specifying algorithms and for reasoning about them. This
is usually done by carefully combining a small number of generic
higher-order functions (such as \textit{map}, \textit{filter}, and
\textit{fold}) that serve as the basic building blocks for writing
high-level programs. The parallelization of algorithms work by
carefully composing an "off-the-shelf" parallel implementation of
each of the building blocks involved in the algorithm. The
underlying parallelization techniques are based on both pipelining
and data parallelism. The essence of this approach is to design a
generic solution once, and to use instances of the design many
times for various applications.

%Figure \caption{An overview of the transformational derivation and the hardware realisation processes.}

In order to develop generic solutions for general parallel
architectures it is necessary to formulate the design within a
concurrency framework such as \textit{CSP} \cite{9,16,125}. Often
parallel functional programs show peculiar behaviors which are
only understandable in the terms of concurrency rather than
relying on hidden implementation details. The formalization in
\textit{CSP} (of the parallel behavior) leads to better
understanding of the described network of processes and allows for
the analysis of its performance. The establishment of refinement
concepts between functional and concurrent behaviors allows for
the generation of parallel implementations for various
architectures. This gives the ability to exploit well-established
functional programming (FP) paradigms and transformation
techniques in order to develop efficient parallel and sequential
\textit{CSP} processes independent from architectural details. The
refinement from functional specification to \textit{CSP}
descriptions is reflected in Figure~\ref{RDM} as transformational
derivation. The transformational derivation is supported by
strategies for parallelism, \textit{CSP} laws, refinement rules
including those for the refinement to \textit{CSP} networks of
processes.

The initial stages of development require a back-end hardware
compiler stage for realizing the developed parallel designs. In
the proposed methodology, \textit{Handel-C} is adopted as the last
stage of development generating the final hardware product. Note
at this point that \textit{Handel-C} language relies on the
parallel constructs in \textit{CSP} to model concurrent hardware
resources. Mostly, algorithms described with \textit{CSP} could be
implemented under \textit{Handel-C}. \textit{Handel-C} enables the
integration with \textit{VHDL} and \textit{EDIF} (Electronic
Design Interchange Format) and thus various synthesis and
place-and-route tools. The \textit{Handel-C} development stage is
described in Figure~\ref{RDM} as an automated compilation step
supported by different code libraries and place-and-route tools
that produces the desired hardware.

The adopted methodology is systematic in the sense that it is
carried out on using step-by-step procedures. The development is
yet manual and applied according to the following informal
procedure:

\begin{itemize}
  \item Specify the algorithm in a functional setting relying on
high-order functions as the main building constructs wherever
necessary.

  \item Apply the predefined set of rules to create the corresponding
\textit{CSP} networks according to a chosen degree of parallelism.

  \item Write the equivalent \textit{Handel-C} code and complete the
hardware compilation.
\end{itemize}

\begin{figure}
	\begin{center}
		\includegraphics [scale = 0.7]%[height=2in,width=2in,angle=0]
		{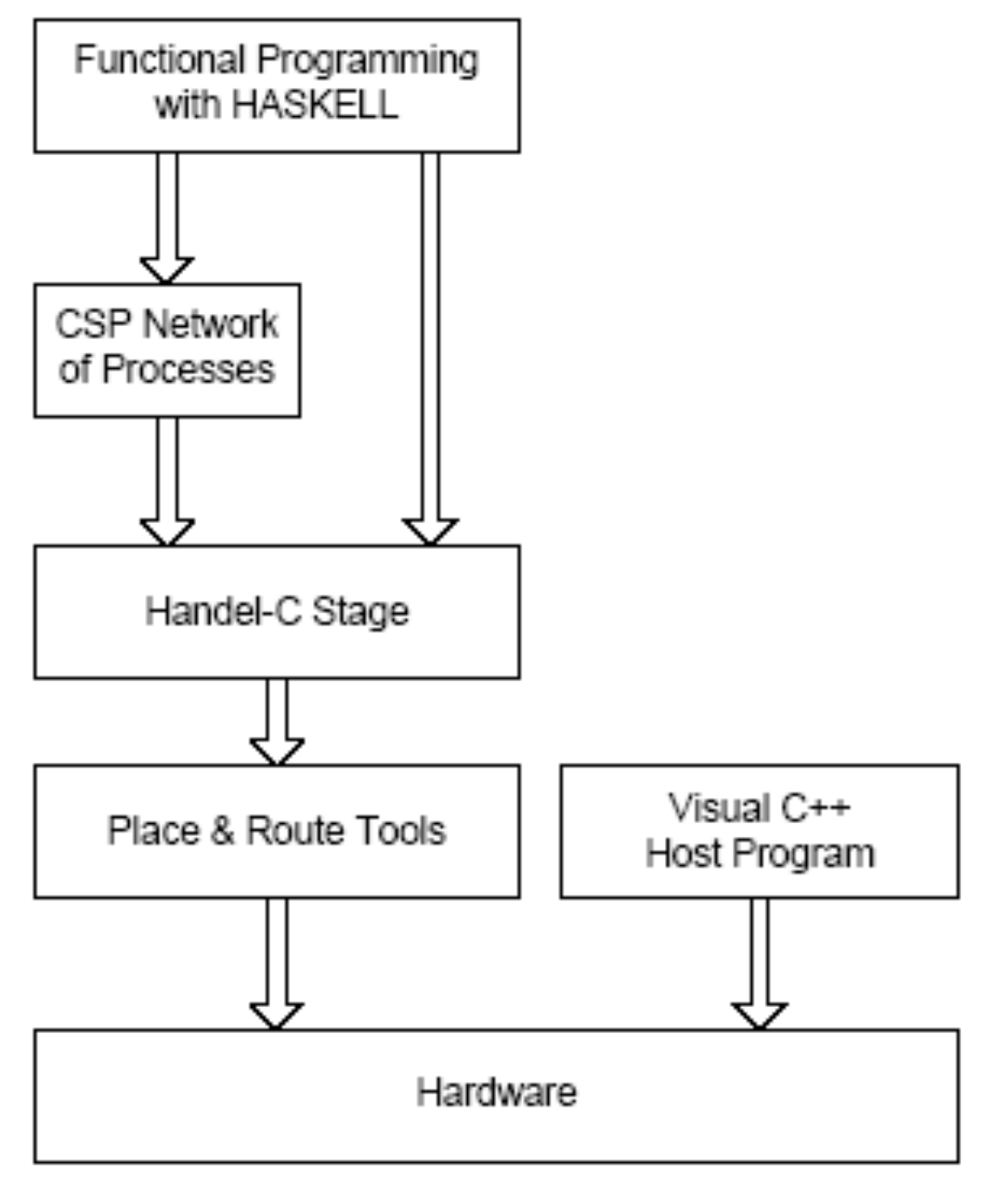}
		\caption{Assisting tools used in the proposed development
			method, including, \textit{Haskell Hugs98} compiler to test
			the specification, \textit{Handel-C} for hardware compilation,
			\textit{Visual C++} integrated development environment to
			create the host program driving the \textit{RC-1000} device
			with its \textit{FPGA}}
		\label{RDMAT}
	\end{center}
\end{figure}

These steps are aided with different compilers and integrated
development environments as shown in Figure~\ref{RDMAT}. The set
of available mathematical rules belong mainly to the refinement to
\textit{CSP} stage. The automation of the development process
including the creation of a preprocessor is currently under
investigation.

The research related to the adopted methodology has been initiated
by Abdallah, investigating the refinement from functional
specifications into concurrency \cite{9}, and presenting a
calculus of decomposition of higher order functions for parallel
programs derivation \cite{8}. Hawkins and Abdallah work included
the formalism for proving the refinement rules for both datatypes
and processes and investigated possible \textit{Handel-C}
implementations \cite{140b}. Case studies where developed for a
\textit{JPEG} decoder, closest pair algorithm, sorting algorithms,
DNA processing algorithms \cite{6,140,143,144}, the
\textit{Kasumi} cryptographic algorithm \cite{d1}, and various
parallel implementations of a matrix multiplication algorithm
\cite{d2}.

The main focus of this paper is on the realization and application
of the theory suggested by the development methodology. An
additional focus is to test the development method and to broaden
its area of use to include an industrial level application.
Furthermore, it includes investigating the performance of the
developed designs by carrying out a thorough analysis and
evaluation. This leads to critically extending, tuning, and
enhancing the suggested method and its realization. In addition,
the current investigation enriches the adopted method by providing
libraries that supports and promotes the method for further
investigation and possibly adoption by mainstream engineers.

The remaining sections of the paper are organized so that
Section~\ref{Background} introduces background material. In
Section~\ref{RW}, related work is discussed. The case study from
cryptography is proposed in Section~\ref{CS}. The analysis and
performance evaluation are included in Sections~\ref{GE} and
~\ref{PA}. Section~\ref{Con} concludes the paper.

%%%%%%%%%%%%%%%%%%%%%%%%%%%%%%%%%%%%%%%%%%%%%%%%%%%%%%%%%%%%%%%%%%%%%%%%%%%%%%%%%%%%%%%%%%%%%%%%%%%%%%

\section{Background} \label{Background}

\textit{Abdallah} and \textit{Hawkins} defined in \cite{140} some
constructs used in the adopted development model. Their
investigation looked in some depth at data refinement; which is
the means of expressing structures in the specification as
communication behavior in the implementation.

The following parts of this section introduces briefly the
proposed steps of development; the functional paradigm,
\textit{CSP}, and \textit{Handel-C}. In addition, we introduce the
basis of the refinement from a functional specification to
networks of \textit{CSP} processes. The benefits of each
development step and the adopted refinement approach are stressed
in Section~\ref{GE}

\subsection{The Functional Paradigm}

Functional programming (\textit{FP}) is quite different from
imperative (or procedural) programming and also from object
oriented programming. \textit{FP}'s main concern is expressions,
where everything reduces to an expression. An expression, that is
a collection of operations and variables, will results in a single
value. Functions are the main building block in functional
programs and could passed around within a program like other
variables. Functional programs are usually \textit{List} oriented,
and they focus description of the problem to be solved rather than
focus on the mechanism of solution. There are currently many
functional languages, one of the most widely used is
\textit{Haskell}\cite{12}.

As a brief overview, we can summarize that functions are
considered as the basic unit of program development and as the
major routes to reuse. In addition, strong typing is considered as
an aid to development pre-implementation, during implementation
and post-implementation. Some of the fundamental features in
\textit{FP} are powerful high-order functions, parametric
polymorphism, the support provided of developing user-defined
datatypes. Other features of no less importance are lazy
evaluation and programming with infinite data structures.
Overloading of function names are not supported in all functional
languages \cite{124}.

High-order functions are an important feature supported in
functional languages. A high-order function is a function which
takes another function as a parameter. The most commonly used
high-order functions are \textit{map}, \textit{zipWith},
\textit{fold}, and \textit{filter}.

The functions \textit{map} and \textit{zipWith} are introduced in
this section. The function \textit{map} takes a list and a
function as parameters, then it applies the input function to all
elements of the input list, for example:

$map$  \ $even$ \ $[1,2,3,4]$ \ $= [False, True, False, True]$

Where \textit{even} is a function that checks wether a number is
even or not.

The function \textit{zipWith} takes two lists and a function as
inputs, then it applies the function on two elements; one element
taken from each input list, for example:

$zipWith$  \ $add$ \ $[1,2,3,4]$ \ $[2,3,4,5]$

$= [3, 5, 7, 9]$

Where \textit{add} is a function that adds two numbers.

Related work adopting \textit{FP} in hardware development is
introduced in Section~\ref{RW}, and the main benefits gained in
using this paradigm in the adopted model are discussed in
Section~\ref{GE}.

\subsection{Communicating Sequential Processes (\textit{CSP})}

The Communicating Sequential Processes (\textit{CSP}) notation is
based on events and processes. A \textit{CSP} process engages in a
series of events, which can be local, or perform channel-based
communication with synchronization capabilities. A channel
communication is an event where at most two processes participate,
one acting as an input and the other as an output.

The alphabets of the components of a concurrent system determine
the overall structure and interface of that system. The fact that
the notation regards basic concurrency operations as primitives
enables the developer to concentrate on the concurrent behavior of
the system without needing to worry about the implementation of
these basic functions.  A synchronization or communication can be
specified in one operation without any concern over how it takes
place \cite{16}.

Valuable features of \textit{CSP} include its strong support for
formal reasoning. \textit{CSP} allows to make generic assertion
about the behavior of the system, such as deadlock freedom.
Moreover, specific assertions of requirements for the behavior of
the process could be done. These assertions are made with
reference to a number of models of the process. Two typical models
are the trace and failures-divergences models.

\textit{CSP} also has the advantage of generality. The primitive
operations of \textit{CSP} are simple enough that almost any form
of concurrency can be represented using them.  Thus, \textit{CSP}
can be used to specify a wide variety of concurrent systems.
Moreover, it can be used to specify the intended functionality of
a message passing system at a formal level without requiring the
system to be modified for a specific architecture (as may be
required by implementation). Employing such features in hardware
development gives the designer the freedom to choose an
appropriate architecture and organization of an implementation
leaving no effect on the original description. Many research
projects have employed \textit{CSP} in hardware design; this is
discussed in Section~\ref{RW}

\subsection{Data Refinement}

In the following, the main concern is explaining the main
constructs and rules to be used in refining a possible functional
specification with its description in \textit{CSP} notation.
Accordingly, we start by presenting some communication entities
used for refining datatypes declared in the initial functional
step of development; these are \textit{Item}, \textit{Stream},
\textit{Vector}, and some of their combined forms. We note here
that the suggested methodology relies on the message passing
technique to implement parallelism.

The \textit{Item} corresponds to a basic type, such as an Integer
data type , and it is to be communicated on a single communicating
channel.

The \textit{Stream} is a purely sequential method of communicating
a list of values (a list is a functional term equivalent to an
array in a language like \textit{C}) . It comprises a sequence of
messages on a channel, with each message representing a value.
Values are communicated one after the other. Assuming the stream
is finite, after the last value has been communicated, the end of
transmission (\textit{EOT}) on a different channel will be
signaled. Given some type \textit{A}, a \textit{Stream} containing
values of type \textit{A} is denoted as $\langle A \rangle$.

Each item to be communicated by the vector will be dealt with in
parallel. A vector refinement of a simple list of items will
communicate the entire structure in a single step. Given some type
\textit{A}, a \textit{Vector} of length \textit{n}, containing
values of type \textit{A}, is denoted as $\lfloor A \rfloor_{n}$.

Whenever dealing with multi-dimensional data structures, for example, lists of lists,
implementation options arise from differing compositions of our primitive data refinements -
streams and vectors. Examples of the combined forms are the Stream of Streams, Streams of Vectors,
Vectors of streams, and Vectors of Vectors. These forms are denoted by: $\langle
S_{1},S_{2},...,S_{n}\rangle$ , $\langle V_{1},V_{2},...,V_{n}\rangle$, $\lfloor
S_{1},S_{2},...,S_{n}\rfloor$, and $\lfloor V_{1},V_{2},...,V_{n}\rfloor$.

\subsection{Process Refinement}

The refinement is continued by looking into the functions
specified in the first stage of development. Accordingly, the
refinement of the formally specified functions to processes is the
key step towards understanding possible parallel behaviour of an
implementation. In this section, the interest is in presenting
refinements of a subset of functions - some of which are
higher-order. A bigger refined set of these functions is discussed
in \cite{9}.

Generally, These highly reusable building blocks can be refined to
\textit{CSP} in different ways. This depends on the setting in
which these functions are used (i.e. with streams, vectors, etc.),
and leads to implementations with different degrees of
parallelism. Note that we don't use \textit{CSP} in a totally
formal way, but we use it in a way that facilitates the later
\textit{Handel-C} coding stage. Recall for the following
subsections that values are communicated through as an
\textit{elements} channel, while a single bit is communicated
through another \textit{eotChannel} channel to signal the end of
transmission in the case of \textit{Streams}.

\subsubsection{Produce}

 The producer process (\textit{PRD}) is fundamental to process refinement. It is used to
produce values on the channels of a certain communication
construct (\textit{Item}, \textit{Stream}, \textit{Vector}, etc.).
These values are to be received and manipulated by another
processes.

\paragraph{Items}

\begin{figure}
	\begin{center}
		\includegraphics [scale=0.6]%[height=2in,width=2in,angle=0]
		{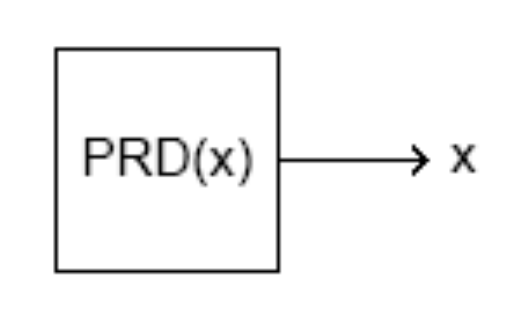}
		\caption{The Produce process (PRD) for items}
		\label{PRDx}
	\end{center}
\end{figure}

 For simple, single item types (\textit{int}, \textit{char}, \textit{bool}, etc.), the
producer process is very simple. This is depicted in
Figure~\ref{PRDx}. Here the output is just a single channel.The
definition in \textit{CSP} notation is very straightforward:

{\small
\[
PRD ~ (Item ~ a) = out.element.channel ~ ! ~ a
\\ \rightarrow SKIP
\]}

% Figure     \caption{The Produce process (PRD) for items}

\paragraph{Streams}

 The producer process for streams is depicted in Figure~\ref{}. As already noted, the
output in this case is a pair of two other channels. One channel
carries the values of the stream, and the other is a simple
channel used to signal \textit{EOT}.

% Figure     \caption{The Produce process (PRD) for streams}
\begin{figure}
	\begin{center}
		\includegraphics [scale=0.6]%[height=2in,width=2in,angle=0]
		{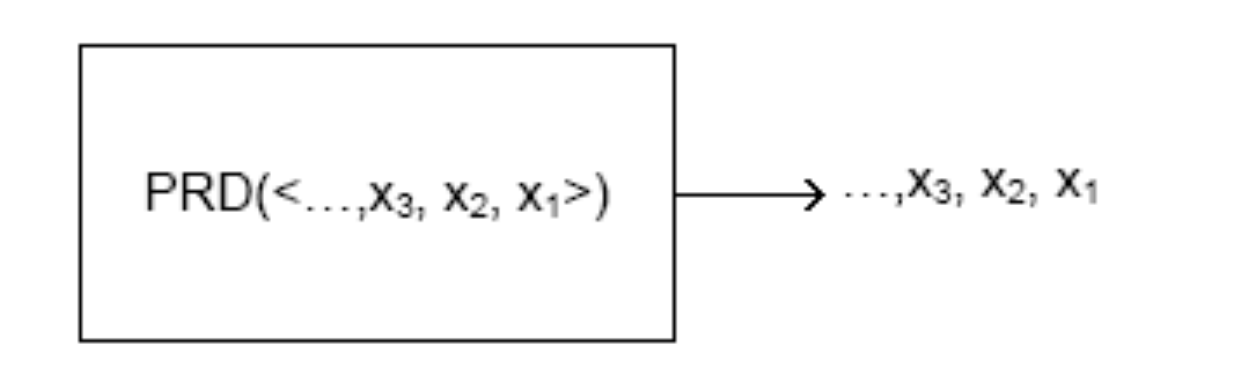}
		\caption{The Producer process (PRD) for streams}
		\label{PRDs}
	\end{center}
\end{figure}

In a more general case, the structure of the values which the stream is carrying is not
necessarily known. These may be simple items, but may also be streams or vectors. Generally,
producing a stream could be described as:

{\small
\begin{equ}
PRD ~ (\stream{s}) =
\\ ((;)_{i=1}^{i=length(s)}
\\ (PRD ~ s_{i}) [out.elements.channel / out]);
\\ out.eotChannel ~ ! ~ eot \rightarrow SKIP
\end{equ}}

This description defines \textit{PRD} as a process that produces
items sequentially (this is described using the sequential
execution operator "\textit{;}"). The number of items is equal to
the length of the stream. After all elements are produced, an end
of transmission signal will be produced on the \textit{eotChannel}
channel.

\paragraph{\normalsize\bf Vectors}

 For vectors of size $n$, $n$ instances of the producer process are composed in parallel, one
for each item in the vector. The output here is an array of channels. This is depicted in
Figure~\ref{PRDv}. A general definition is given below:

{\small
\begin{equ}
PRD ~ (\lfloor v \rfloor_{n}) = \interleave_{i=1}^{i=n} \\ (PRD ~ v_{i}) [
out.elements_{i}.channel / out ]
\end{equ}}

The operator $\interleave_{i=1}^{i=n}$ is used to indicate that
$n$ copies of the process $PRD ~ v$ for producing items will be
running concurrently. \textit{PRD} is described as a processes
that runs concurrently $n$ instances ( of a processes that
produces single items).

% Figure \caption{The Produce process (PRD) for vectors}
\begin{figure}
	\begin{center}
		\includegraphics [scale= 0.5]%[height=2in,width=2in,angle=0]
		{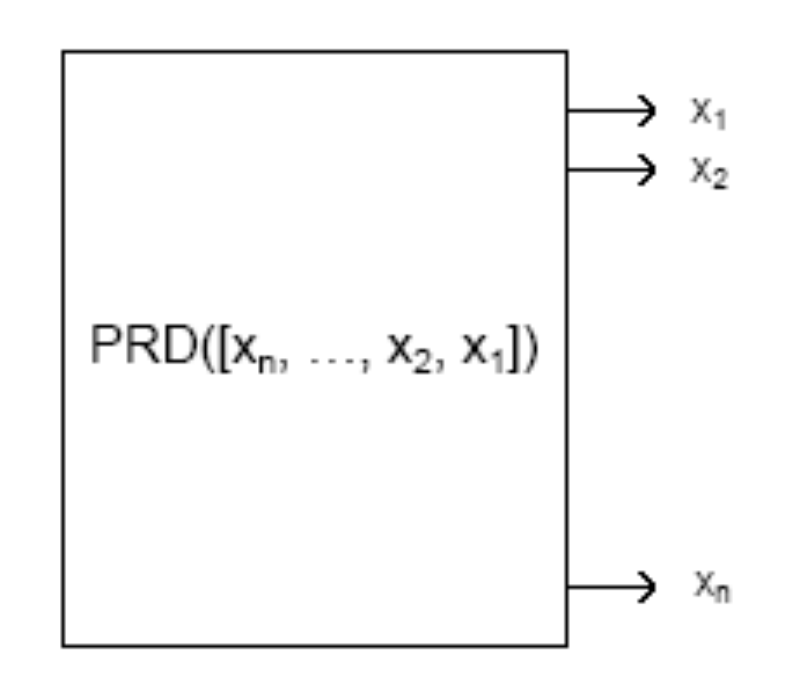}
		\caption{The Producer process (PRD) for vectors}
		\label{PRDv}
	\end{center}
\end{figure}

A process \textit{STORE} stores a communication construct in a
variable. We use this process to store items, vectors, streams, or
combinations of vectors and streams. A subscript letter is used
with the processes \textit{PRD} and \textit{STORE} to indicate the
type of communication. We sometimes omit this subscript if the
communication structure is clear from context.

\subsubsection{Feeding Processes}

 The feed operator in \textit{CSP} models function application. The feed operator is written
$\rhd$. The feed operator takes two processes, composes them together in parallel, and renames
both the output of the first and the input of the second to a new name, which is then hidden.
Given the lifted concepts of \textit{CSP} channel renaming and hiding, the definition can remain
the same regardless of the type of the communicating construct (\textit{Item}, \textit{Stream},
\textit{Vector} or any combination).

{\small
\[
P \rhd Q = \\ (P [mid / out] ~ || ~ Q [mid / in]) \verb"\" \{ mid \}
\]
}

\subsubsection{Formal Process Refinement}

 Given the definition of a feed operator that operates on processes, a formal definition of
process refinement could be delivered. Consider a function $f$,
which takes input values of type $A$ and returns values of type
$B$. Assume that the data refinement step has already been
performed, such that $A$ and $B$ are both types of some
transmission value:

{\small
\[
f :: A \rightarrow B
\]
}

Then, consider a potential refinement for a function $f$, a
process $F$. The operator $\sqsubseteq$ denotes a process
refinement, where the left hand side is a function, and the right
hand side is a process. To state that $f$ is refined to $F$, or in
other words, the process $F$ is a valid refinement of the function
$f$, the following may be used:

\[
f \sqsubseteq F
\]

The rules of refinement were proven once in \cite{9} and applied
in this paper refine a functional specification into a network of
communicating processes.

%%%%%%%%%%%%%%%%%%%%%%%%%%%%%%%%%%%%%%%%%%%%%%%%%

\subsubsection{MAP the Process Refinement of the Higher-order Function \textit{map}}

 Now the attention is turned to the refinement of the widely used higher-order function
\textit{map} \cite{140} . Employing this function in stream and vector settings is presented. The
refinement for combined structures is to be made in a similar way.

\paragraph{Streams}

\begin{figure}
	\begin{center}
		\includegraphics [scale=0.6]%[height=2in,width=2in,angle=0]
		{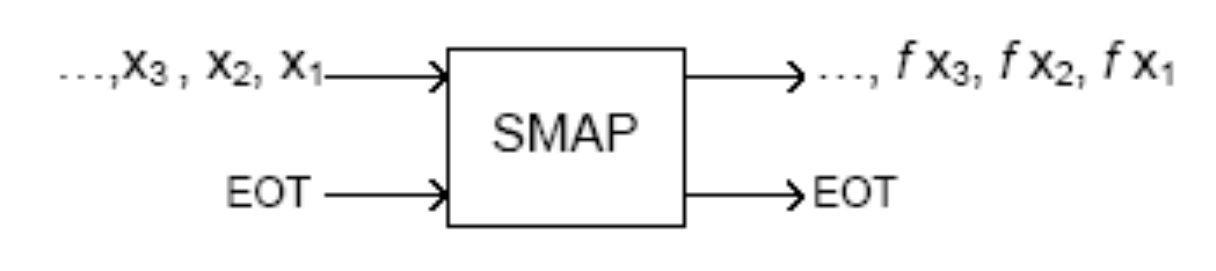}
		\caption{The SMAP process for streams}
		\label{SMAP}
	\end{center}
\end{figure}

A process implementing the functionality of $map ~ f$ in stream terms should input a stream of
values, and output a stream of values with the function $f$ applied (See Figure~\ref{SMAP}).

In general, the handling of the \textit{EOT} channels will be the same. However, the handling of
the value will vary depending on the type of the elements of the input and output stream.

{\small
\[
SMAP(F) =\\
\mu  X ~ \bullet  in.eotChannel ~ ? ~ eot \rightarrow \\ out.eotChannel ~ ! ~ eot \rightarrow SKIP \\
                   \choice \\
                   F [in.elements.channel / in, \\ out.elements.channel / out] ; X
\]
}

%Figure \caption{The SMAP process for streams}

\paragraph{Vectors}

\begin{figure}
	\begin{center}
		\includegraphics [scale = 0.5]
		{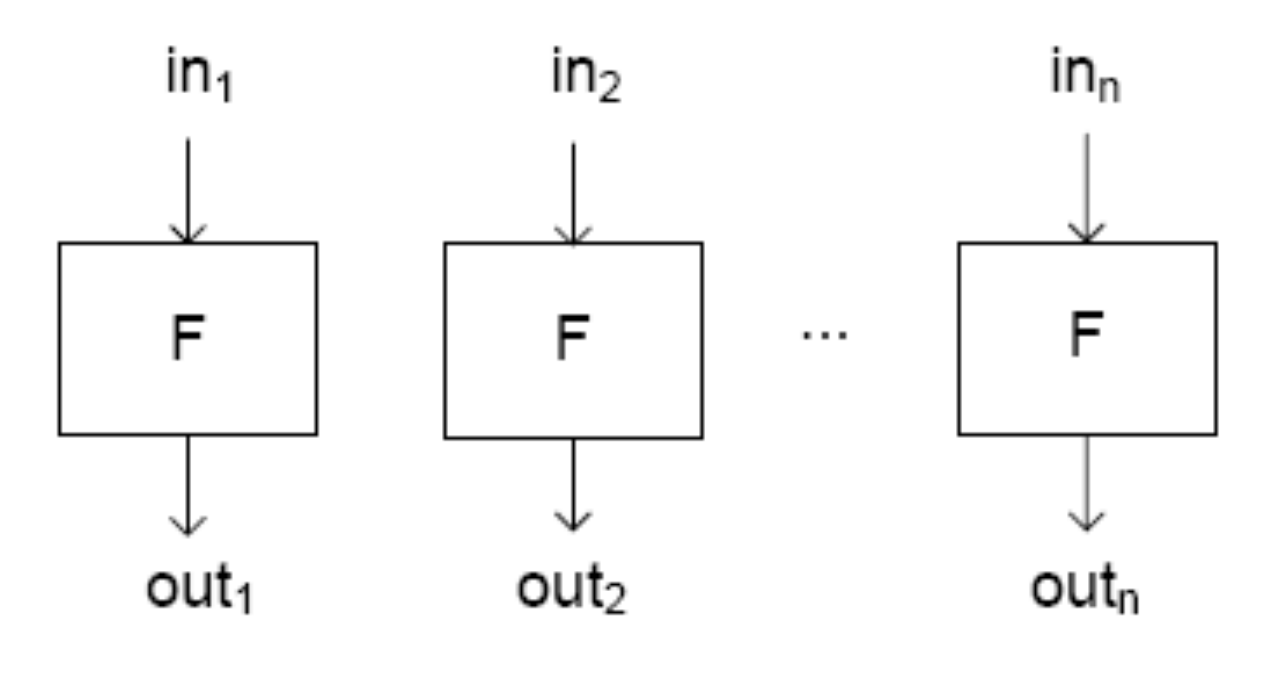}
		\caption{The VMAP process for vectors}
		\label{VMap}
	\end{center}
\end{figure}

 In functional terms, the functionality of $map ~ f$ in a list setting is modelled by $vmap
~ f$ in the vector setting. Consider $F$ as a valid refinement of the function $f$. The
implementation of $VMAP$ can then proceed by composing $n$ instances of $F$ in parallel, and
directing an item from the input vector to each instance for processing (See Figure~\ref{VMap}).
In \textit{CSP} we have:

{\small
\begin{equ}
VMAP_{n}(F) & = & \\
\interleave_{i=1}^{i=n} F [ in_{i} / in,
out_{i} / out]
\end{equ}
}

%Figure \caption{The VMAP process for vectors}

%%%%%%%%%%%%%%%%%%%%%%%%%%%%%%%%%%%%%%%%%%%%%%%%

\subsubsection{ZIPWITH the Process Refinement of the Higher-order Function \textit{zipWith}}

 Recall another higher-order function, namely \textit{zipWith}. This function is used to zip
two lists (taking one element from each list) with a certain operation. Formally:

{\small
\[
zipWith :: \\ (A\rightarrow B\rightarrow C) \rightarrow [A]\rightarrow[B]\rightarrow[C]
\]
\[
zipWith ~ (\oplus) ~ [x_{1}, x_{2}, ... x_{n}] [y_{1}, y_{2}, ... y_{n}] = \\ [x_{1} \oplus y_{1},
x_{2} \oplus y_{2},..., x_{n} \oplus y_{n}]
\]
}

\begin{figure}
	\begin{center}
		\includegraphics [scale =0.6]
		{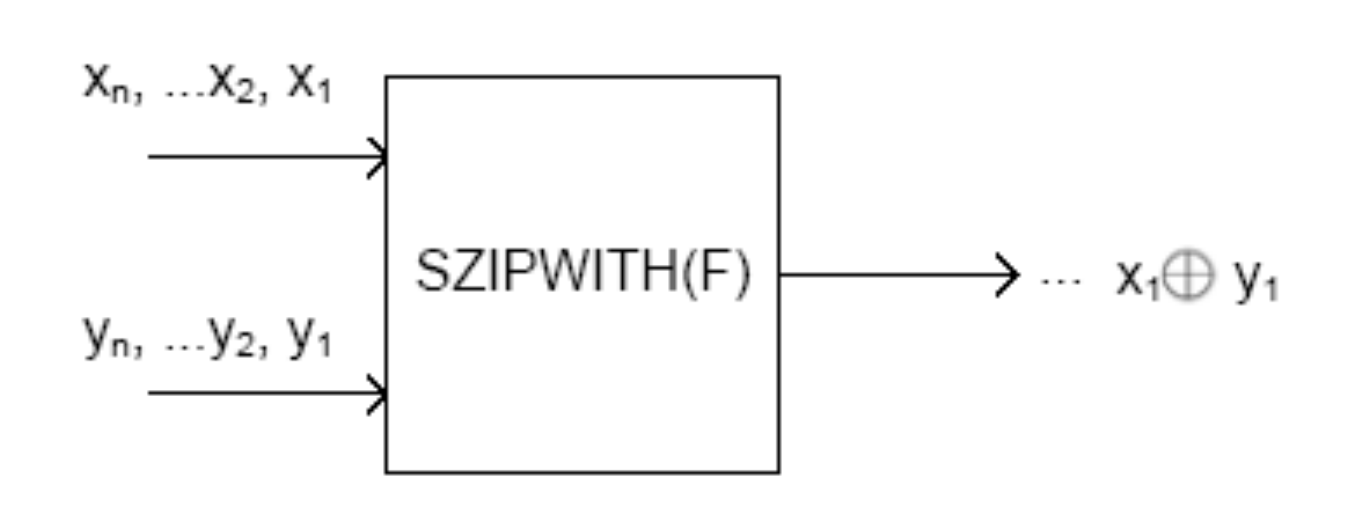}
		\caption{The SZIPWITH process for streams}
		\label{SZipWith}
	\end{center}
\end{figure}

\paragraph{Streams}
The process implementation of ($zipWith ~ f)$ in stream terms should input two streams of values,
and output a stream of values with the function $f$ applied (See Figure~\ref{SZipWith}).

% Figure     \caption{The SZIPWITH process for streams}

Again, the handling of the \textit{EOT} channel will be the same. Nevertheless, the handling of
the value will vary depending on the type of the input and output streams elements.

{\small
\[
\begin{array}{lcl}
SZIPWITH(F)  =  \\
\mu  X ~ \bullet  in.eotChannel ~ ? ~ eot \rightarrow \\ out.eotChannel ~ ! ~ eot \rightarrow SKIP \\
\choice \\
F [in_{1}.elements.channel / in_{1}, \\ in_{2}.elements.channel /
in_{2}, \\
out.elements.channel / out] ; X
\end{array}
\]
}

\begin{figure}
	\begin{center}
		\includegraphics [scale = 0.6]
		{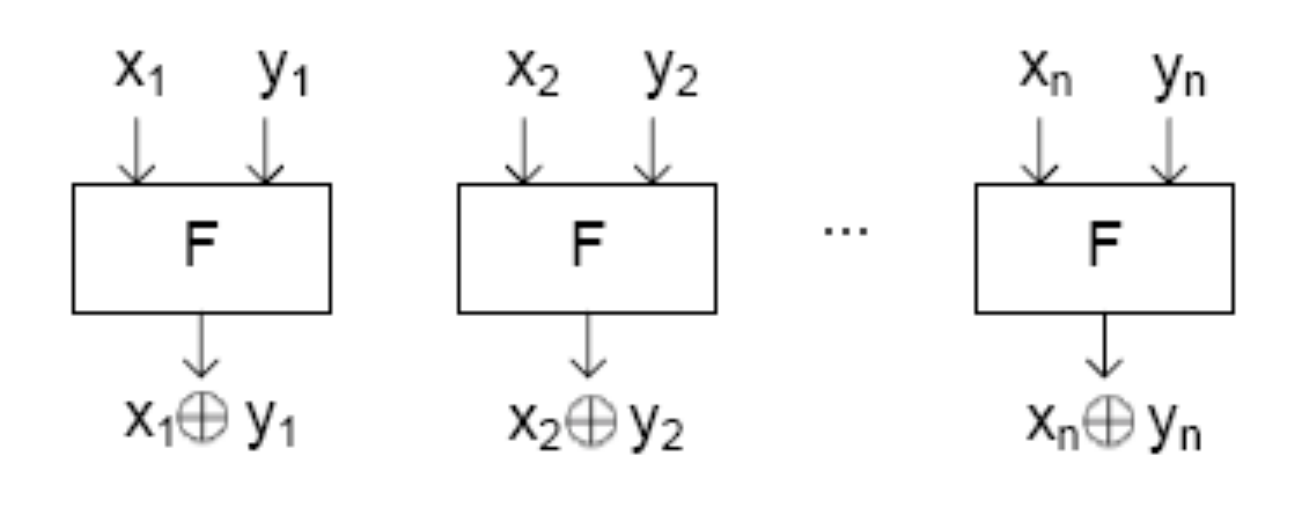}
		\caption{The VZIPWITH process for vectors}
		\label{VZipWith}
	\end{center}
\end{figure}

\paragraph{Vectors} To implement the data parallel version of this
higher-order function, we refine it to a process \textit{VZIPWITH}
that takes two vectors as input and zips the two lists with a
process \textit{F}; \textit{F} is a refined process from the
function ($\oplus$). This is depicted in Figure~\ref{VZipWith}.

{\small
\[
vzipWith ~ (\oplus):: \lfloor A \rfloor_{n} \rightarrow ,\lfloor B \rfloor_{n} \rightarrow \lfloor
C \rfloor_{n}
\]

\[
VZIPWITH ~ (\oplus) = \\ \interleave_{i=1}^{i=n}F[out_{i}/out,c_{i}/in_{1},d_{i}/in_{2}]
\]
}

%Figure     \caption{The VZIPWITH process for vectors}

\subsection{Handel-C as a Stage in the Development Model}

Based on datatype refinement and the skeleton afforded by process
refinement, the desired reconfigurable circuits are built. Circuit
realisation is done using \textit{Handel-C}, as it is based on the
theories of \textit{CSP} \cite{16} and \textit{Occam} \cite{B3}.

From a practical standpoint, each refined datatype is defined as a
structure in \textit{Handel-C}, while each process is implemented
as a \textit{macro} \textit{procedure}. We divide the constructs
corresponding to the \textit{CSP} stage into two main categories
for organisation purposes. The first category represents the
definitions of the refined datatypes. The second category
implements the refined processes. The refined processes are
divided into different groups. The \textit{utility}
\textit{processes} group contains macros responsible for
producing, storing, sinking, broadcasting data, etc. The
\textit{basic} \textit{processes} group contains macros that
correspond to simple arithmetic and logical operations. These
basic processes could be simple addition, multiplication, etc. The
\textit{higher-order} \textit{processes} group contains the macros
realising the \textit{CSP} implementations corresponding to the
higher-order functions. A separate group contains the macros that
handle the \textit{FPGA} card setup and general functionality. The
reusable macros found in these groups serves as building blocks
used for constructing a certain specified algorithm.

\subsubsection{Datatypes Definitions}

 The datatypes definitions are implemented using structures. This method supports recursive
as well as simple types. The definition for an \textit{Item} of a type \textit{Msgtype} is a
structure that contains a communicating channel of that type.

{\small
\begin{verbatim}
 #define Item(Name, Msgtype)
    struct {
        chan Msgtype    channel;
        Msgtype         message;
        } Name
\end{verbatim}}

For generality in implementing processes the type of the communicating structure is to be
determined at compile time. This is done using the \textit{typeof} type operator, which allows the
type of an object to be determined at compile time. For this reason, in each structure we declare
a \textit{message} variable of type \textit{Msgtype}.

A stream of items, called \textit{StreamOfItems}, is a structure with three declarations a
communicating channel, an \textit{EOT} channel, and a \textit{message} variable \cite{140}:

{\small
\begin{verbatim}
 #define StreamOfItems(Name, Msgtype)
    struct {
        Msgtype         message;
        chan Msgtype    channel;
        chan Bool       eotChannel;
        } Name
\end{verbatim}}

A vector of items, called \textit{VectorOfItems}, is a structure with a variable \textit{message}
and another array of sub-structure elements \cite{140}.

{\small
\begin{verbatim}
 #define VectorOfItems(Name, n, Msgtype)
    struct {
        struct {
            chan Msgtype    channel;
            } elements[n];
        Msgtype     message;
        } Name
\end{verbatim}}

Other definitions are possible, but it affects the way a channel is called using the structure
member operator (.). Examples of different extended definitions are as follows (the first
definition reuses the \textit{Item} structure, while the second one employs channel arrays
supported in \textit{Handel-C}):

{\small
\begin{verbatim}
 #define VectorOfItems(Name, n, Msgtype)
    struct {
        struct {
            Item(element, MsgType);
            } elements[n];
        } Name

 #define VectorOfItems(Name, n, Msgtype)
    struct {
        chan Msgtype    channel[n];
        Msgtype         messages;
        } Name
\end{verbatim}}

\subsubsection{Utilities Macros}
The utility processes used in the implementation are related to
the employed datatypes. The \textit{Handel-C} implementation of
these processes relies on their corresponding \textit{CSP}
implementation. An instance of these utility macros is shown in
the following code segment:

{\small
\begin{verbatim}
 macro proc ProduceItem(Item, x){
    Item.channel ! x;}

 macro proc StoreItem(Item, x){
    Item.channel ? x;}
\end{verbatim}}

\subsubsection{Higher-Order Processes Macros}

 An example for an implementation in \textit{Handel-C} of the \textit{CSP} refinement of a
higher-order function (\textit{map}) is done as follows. The
process runs through a loop which terminates when the variable
\textit{eot} is set to true. At each step of the loop, the process
enters a wait state until either the \textit{EOT} or the value
channel of the input stream is willing to communicate. If the
\textit{EOT} channel is willing to communicate, the input is
consumed from it and stored in the variable \textit{eot}, then
output an \textit{EOT} message for the output stream.  If the
value channel of the input stream is willing to communicate, the
value is consumed then $F$ is applied to it giving the result on
the output stream channel.

{\small
\begin{verbatim}
 macro proc
  SMAP (streamin, streamout, F){
   Bool eot;
   eot  = False;
   do{
     prialt{
     case streamin.eotChannel ? eot:
        streamout.eotChannel ! True;
        break;
     default:
        F(streamin.elements,
           streamout.elements);
         break;
      }} while (!eot)}
\end{verbatim}}

We turn the attention to providing a definition in
\textit{Handel-C} for the behaviour of the process \textit{VMAP}.
Here we can employ \textit{Handel-C}'s enumerated \textit{par}
construct to place $n$ instances of the process $F$ in parallel.
Each instance is passed to the corresponding channels from both
the input and output channels.

{\small
\begin{verbatim}
 macro proc
  VMAP (n, vectorin, vectorout, F) {
   typeof (n) c;
   par (c = 0 ; c < n ; c++){
      F(vectorin.elements[c],
        vectorout.elements[c]);}}
\end{verbatim}}

\subsection{Evaluation Tools and Performance Metrics}

Different tools are used to measure the performance metrics used
for the analysis. These tools include the design suite
(\textit{DK}) from \textit{Celoxica}, where we get the number of
\textit{NAND} gates for the design as compiled to (\textit{EDIF}).
The \textit{DK} also affords the number of cycles taken by a
design using its simulator. Accordingly, the speed of a design
could be calculated depending on the expected maximum frequency of
the design.

To get the practical execution time as observed from the host computer, the \textit{C++}
high-precision performance counter is used. The counter probes the execution of the design after
loading the image of the design into the \textit{FPGA} till termination. Practically, the
probation comes directly after writing a control signal to the \textit{FPGA} enabling execution.
The counter stops immediately after receiving a signal through reading the status register.
According to this measurement the speed of execution is calculated.

The information about the hardware area occupied by a design, i.e.
number of Slices used after placing and routing the compiled code,
is determined by the \textit{ISE} place and route tool. In the
current investigation the only used metrics are the number of
Slices and the Total Equivalent Gate Count for a design.

\section{Related Work} \label{RW}
In this section we define four perspectives, not necessarily
mutually exclusive or unconnected, to be considered for relating
our work with its global literature:

\begin{itemize}
  \item Purpose: Related to frameworks created for refining correct hardware implementations.

  \item Implementation Framework: Related to the use of the Functional Paradigm in hardware
  development. Related work in this area might also meet the purpose of developing correct
  reconfigurable hardware.

  \item Description: Related to the use of \textit{CSP} in hardware development.

  \item Application: Related to the use of \textit{FPGAs} in implementing the Serpent cryptographic algorithm.
\end{itemize}

The idea for deriving implementations from the specification
through correct well defined refinement steps has been motivated
by many technical facts. For instance, the limitations in commonly
used synthesis tools and formal verification techniques utilized
in equivalence checking between the synthesized hardware and the
abstract specification \cite{r19}. Many frameworks for developing
correct hardware has been brought out in the literature
\cite{r16,r17,r18,r19}. Our work meets these multi-stage
frameworks in their aim of refining correct hardware from
specification.

The Provably Correct Systems project (\textit{ProCoS}) suggested a
mathematical basis for the development of embedded and real-time
computer systems. They used \textit{FPGAs} as a back-end hardware
for realizing their developed designs \cite{r16}.

In \cite{r18}, a formal approach to correctly generate an
architecture-level model of a system from its specification model
is proposed . The proposed approach relies on formal
transformations to refine a specification model into a provably
correct architectural model. Tools have been created to support
automatic generation of refined models \cite{r19}.

The attractions for using the functional paradigm in hardware
development incited many researchers. This triggered many
investigations in this area, such as \textit{Lava} \cite{122},
\textit{Hawk} \cite{r1,r2}, \textit{Hydra} \cite{r4}, \textit{HML}
\cite{r7}, \textit{MHDL} \cite{r6}, \textit{DDD} system \cite{r8},
\textit{SAFL} \cite{r10}, \textit{MuFP} \cite{r11}, \textit{Ruby}
\cite{r12}, and \textit{Form} \cite{r14}.

The compiled \textit{Occam} into \textit{FPGAs} \cite{108}
\cite{r21} and the \textit{Handel-C} compiler \cite{129} are
considered as the major work introducing \textit{CSP} in hardware
development. Susan Stepney at the University of York \cite{125}
\cite{r20} investigated ways to translation between \textit{CSP}
and \textit{Handel-C}. \textit{Handel-C} compiler is used to map
designs onto \textit{FPGAs}. The suggested translation uses
\textit{FDR2} as a front-end specification and proof tool, then
automatically translates the formal designs into executable
\textit{Handel-C}.

Many efforts have been put to efficiently implement the
\textit{Serpent} in hardware. R. Anderson proposed in \cite{s6}
the \textit{Serpent} algorithm and evaluated its performance under
different processing systems. Adam et al in \cite{s2} presented an
\textit{FPGA} implementation and performance evaluation of the
\textit{Serpent}. Multiple architecture options of the
\textit{Serpent} algorithm were explored with a strong focus being
placed on high-speed implementations. Bora in \cite{s8}
investigated the possibilities of realising the \textit{Serpent}
using \textit{FLEX10K} \textit{ALTERA} \textit{FPGAs} series. The
implementations of this algorithm was introduced in \cite{s7} with
an effort to determine the most suitable candidate for hardware
implementation within commercially available \textit{FPGAs}.

%%%%%%%%%%%%%%%%%%%%%%%%%%%%%%%%%%%%%%%%%%%%%%%%%%%%%%%%%%%%%%%%%%%%%%%%%%%%%%%%%%%%%%%%%%%%%%%%%%%%%%

\section{Case Study: The Serpent Cryptographic Algorithm} \label{CS}

The \textit{Serpent} algorithm is chosen as a test case for the
proposed development model. The motivation behind choosing the
\textit{Serpent} is its proven strength and suitability for
hardware implementation \cite{s6}. The \textit{Serpent} algorithm
is a 32-round substitution-permutation (\textit{SP}) network
operating on four 32-bit words. The algorithm encrypts and
decrypts 128-bit input data and a key of 128, 192, or 256 bits in
length. The \textit{Serpent} algorithm consists of three main
blocks an initial permutation (\textit{IP}), A 32-round block, and
a final Permutation (\textit{FP}). One round function is comprised
of three operations occurring in sequence. These are bit-wise XOR
with the 128-bit round key, substitution via 32 copies of one of
eight S-boxes, and data mixing via a linear transformation. These
operations are performed in each of the 32 rounds with the
exception of the last round. In the last round, the linear
transformation is replaced with a bit-wise XOR with a final
128-bit key.

This section develops parallel implementations of the
\textit{Serpent} algorithms showing all stages of development and
the results of testing. The following subsections presents the
functional specification, followed by the refinement and the
implementation in \textit{Handel-C}. Various designs with
different degrees of parallelism are investigated. Different
solutions are presented to some realization pitfalls. The final
section presents the results of running the compiled designs with
comparison among different processing systems.

\subsection{Formal Functional Specification} Two main building blocks construct the
\textit{Serpent}, the key scheduling block and the encryption (decryption) block. The key
scheduling block inputs the private key and outputs the desired 132 subkeys. The encryption block
inputs data segments representing the plaintext and outputs the corresponding ciphered data
segments. The formal functional specification employs the following names used for clarifying
types definitions.

 {\small \begin{verbatim}
 type Private   = [Bool]
 type SubKey    = [Bool]
 type DataBlock = [Bool]
 \end{verbatim}}

The following subsections present the specification of the
\textit{Serpent} algorithm. The implementation of the
specification under \textit{HUGs98} \textit{Haskell} compiler is
tested at the unit, component and integration levels.

%%%%%%%%%%%%%%%%%%%%%%%%%%%%%%%%%%%%%%%%%%%%%%%%%%%%%%%%%%%%%%%%%%%%%%%%%%%%%%%%%

\subsubsection{Key Scheduling}
Two main steps are carried out to generate the required 132 32-bit subkeys for the
\textit{Serpent}. The algorithm for generation is as follows:

\begin{itemize}
  \item Generate an intermediate list \textit{ws} by:
        \begin{itemize}
        \item Padding the input key to 256-bit if necessary.
        \item Then, partitioning the key into eight segments of equal length (32-bit) $ws_{0},.., ws_{7}$.
        \item Then, expanding these to intermediate prekeys $ws_{8},.., ws_{139}$ by the following recurrence:

        $ws_{i} := (ws_{i-8} \oplus ws_{i-5} \oplus ws_{i-3} \oplus ws \oplus  9e3779b9_{hex} \oplus (i-8))<<_{11}$

        where ($<<_{n}$) is the n-element left circular shift operator.
        \end{itemize}

  \item The round subkeys \textit{ks} are now calculated from the prekeys \textit{ws} using the
  S-boxes as follows:
\\

{\small
  $\{k_{0}; k_{1}; k_{2}; k_{3}\} = S3(w_{0}; w_{1}; w_{2}; w_{3})$
  $\{k_{4}; k_{5}; k_{6}; k_{7}\} = S2(w_{4}; w_{5}; w_{6}; w_{7})$
  $\{k_{8}; k_{9}; k_{10}; k_{11}\}= S1(w_{8}; w_{9}; w_{10}; w_{11})$
  $\{k_{12}; k_{13}; k_{14};k_{15}\} = S0(w_{12}; w_{13}; w_{14}; w_{15})$
  $\{k_{16}; k_{17}; k_{18}; k_{19}\}$ \ \ \ \ $= S7(w_{16}; w_{17}; w_{18}; w_{19})$
  \\

  $...$
  \\

  $\{k_{124}; k_{125};k_{126}; k_{127}\} = S4(w_{124}; w_{125}; w_{126}; w_{127})$
  \\

  $\{k_{128}; k_{129}; k_{130}; k_{131}\} = S3(w_{128}; w_{129}; w_{130}; w_{131})$
}

\end{itemize}

\begin{figure}
	\begin{center}
		\includegraphics [scale=0.6] %[height=2in,width=2in,angle=0]
		{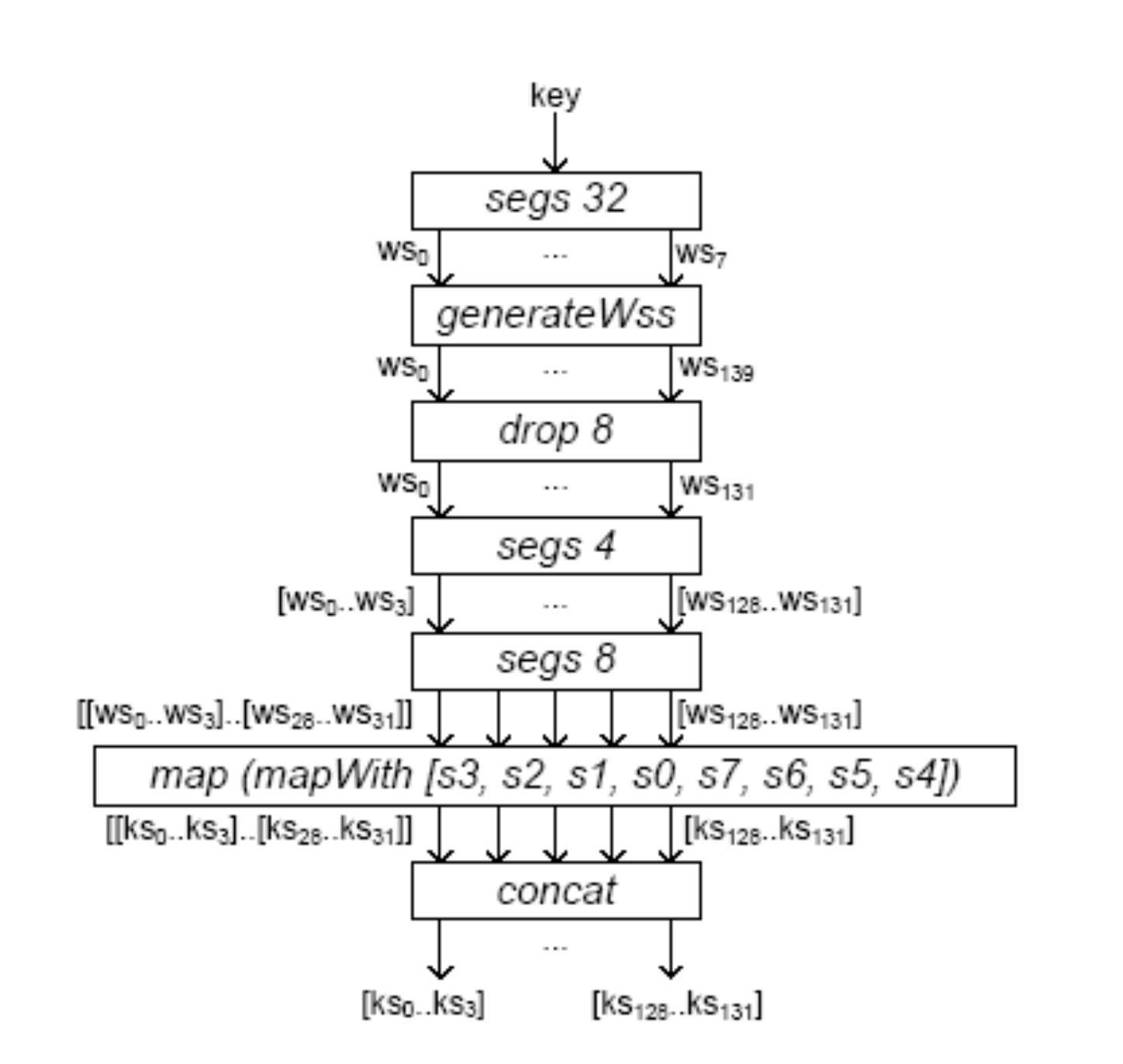}
		\caption{Steps for \textit{Serpent} subkeys generation}
		\label{KeySchSteps}
	\end{center}
\end{figure}

The function \textit{keySchedule} formally specifies the above
algorithm. This function inputs the private key and outputs the
desired subkeys following the steps clarified in
Figure~\ref{KeySchSteps}. This figure also shows the format of the
final output as ordered for later use in the functions specifying
the encryption.

 {\small \begin{verbatim}
 keySchedule :: Private -> [[SubKey]]
 keySchedule key = concat kss
 where
 ws = drop 8 (generateWs 8 (segs 32 key))
 kss = map (mapWith [s3, s2, s1, s0,
                      s7, s6, s5, s4])
                  (segs 8 (segs 4 ws))
\end{verbatim}}

The application of the S-boxes is done by mapping the function
\textit{(mapWith [s3, s2, s1, s0, s7, s6, s5, s4])} over the
prepared segmentation of \textit{ws} \textit{(segs 8 (segs 4
ws))}. Note that the length of the list \textit{ws} at this point
is 132 elements. Grouping this list into segments of four and then
of eight, will give four lists each of eight 4-element sublists,
covering 128 elements from \textit{ws}. The remaining 4 elements
constitutes a final list of four elements. With the lazy
evaluation property found in functional programming, the final
mapped \textit{mapWith} only applies the function \textit{s3} to
the remaining list. This will give the desired output list of
lists representing the 132 round subkeys.

%Figure \caption{Steps for \textit{Serpent} subkeys generation}

The \textit{generateWs} responsible for generating the prekeys is specified as follows:

 {\small \begin{verbatim}
 generateWs :: Int -> [[Bool]] -> [[Bool]]
 generateWs i ws
   | ((i < 140) && (i > 7)) =
          (generateWs (i+1) (ws ++ [wsD]))
   | otherwise = ws
   where
    wsD = (shift 11 (foldr1 fullexor
    [(ws!!(i-8)), (ws!!(i-5)),
    (ws!!(i-3)), (ws!!(i-1)),
    const, (itob (i-8))]))

    const = concat
    (map itob.htoi ["9e37", "79b9"])
\end{verbatim}}

The S-boxes are specified using the logic functions
\textit{fullexor}, \textit{fullOR}, \textit{fullAND}, and
\textit{fullComplement}. These corresponds to the full-word
bitwise version of \textit{XOR}, \textit{OR}, \textit{AND}, and
\textit{NOT} logic operations. For instance, the first S-box is
specified as the function \textit{s0} with a list of list of
\textit{bool} as input and output. The input list elements
\textit{[a, b, c, d]} are distributed to different operations
computing for the final output list \textit{[w, x, y, z]}.
Temporary variables used to compute the final output list are
grouped to be zipped with their operation using the higher-order
function \textit{zipWith} . The current specification does not
reflect the order that these operations should be carried out. A
dependency analysis has to be done aiding the later refinement.
Note that the decryption inverse S-boxes are specified in a
similar way. In the following we show the specification of the
\textit{s0} function.

 {\small \begin{verbatim}
 s0 :: [[Bool]] -> [[Bool]]
 s0 [a,b,c,d] = [w, x, y, z]
   where
   [t01, t03, z, t06, y,
    t12, t13, t15, t17, x ]=
    zipWith
       fullexor [b, a, t02, a, t09,
                c, t07, t06, w, t12]

                [c, b, t01, d, t08,
                d, t11, t13, t14, t17]

   [t05,t07, t02] =
    zipWith fullOR [c, b, a] [z, c, d]

   [t08, t09, t11, t14] =
    zipWith fullAND [d, t03, t09, b]
                    [t05, t07, y, t06]

   w = fullComplement t15
\end{verbatim}}

%%%%%%%%%%%%%%%%%%%%%%%%%%%%%%%%%%%%%%%%%%%%%%%%%%%%%%%%%%%%%%%%%%%%%%%%%%%%%%%%%

\subsubsection{Serpent Block Cipher}
\begin{figure}
	\begin{center}
		\includegraphics [scale=0.55] %[height=2in,width=2in,angle=0]
		{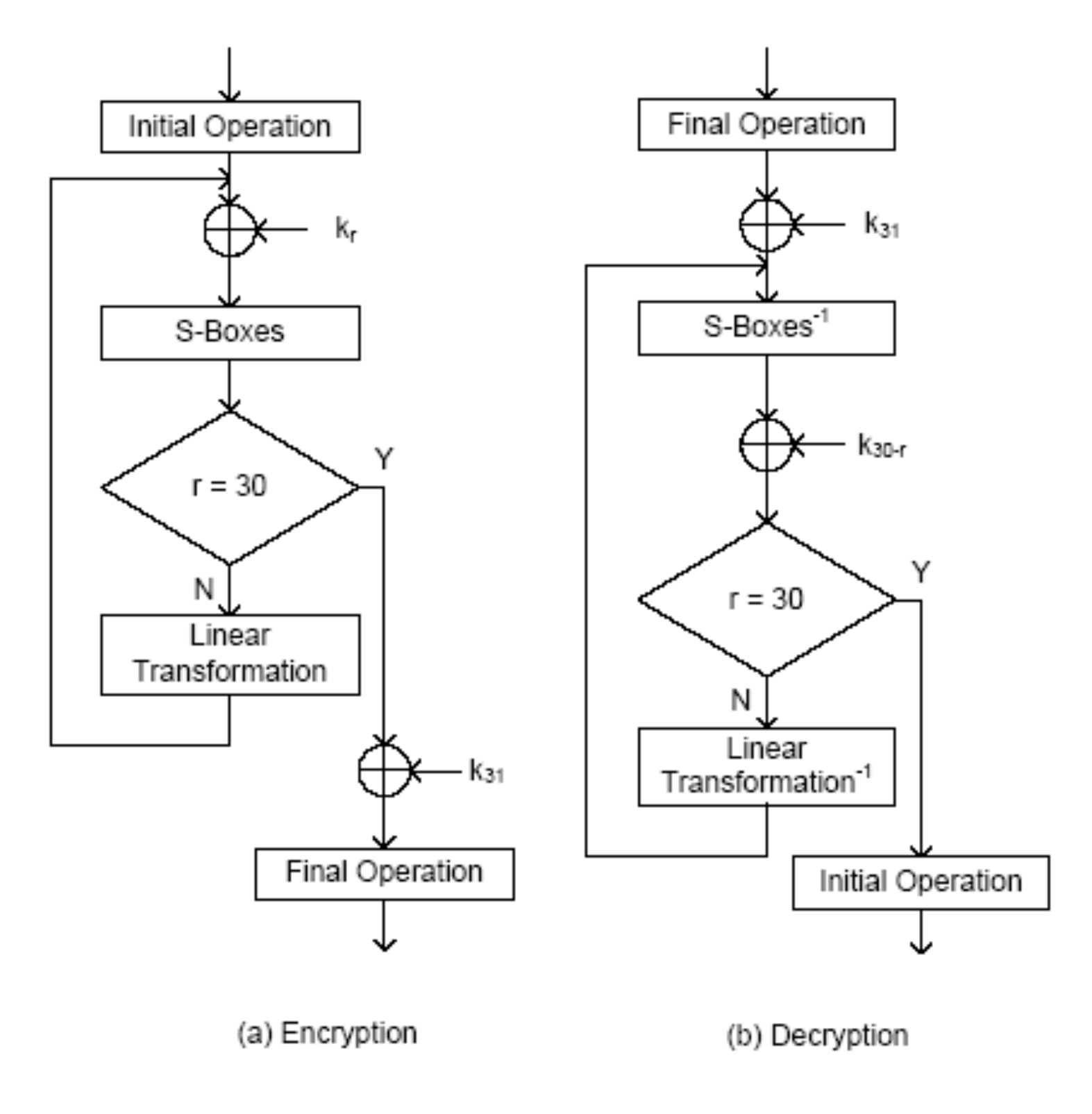}
		\caption{Serpent encryption (a) and decryption (b) flowcharts}
		\label{EncDec}
	\end{center}
\end{figure}

Flowcharts showing the steps to carry out the encryption and the
decryption are shown in Figure~\ref{EncDec}. Decryption is
different from encryption in that the inverse of the S-boxes must
be used in the reverse order, as well as the inverse linear
transformation and reverse order of the subkeys.

%Figure     \caption{Serpent encryption (a) and decryption (b) flowcharts}

A functional specification formulates \textit{Serpent} encryption
as a function \textit{serpentEncrypt}. This function works by
firstly inputting a list of lists of data blocks. Then, it maps
the function \textit{serpentEncryptSeg}, responsible for a single
128-bit data block encryption, with the input private key to all
the input list elements. The functional specification of
\textit{serpentEncrypt} is as follows:

 {\small \begin{verbatim}
 serpentEncrypt :: [[DataBlock]] -> Private
                   -> [[DataBlock]]

 serpentEncrypt inputs key =
 map (serpentEncryptSeg(keySchedule key))
    inputs
\end{verbatim}}

The formalised function \textit{serpentEncryptSeg} inputs the
generated round subkeys in a form of a list of lists, besides, the
128-bit plaintext input data block. The first 31 rounds subkeys
are taken from the input list of subkeys and zipped in a list of
pairs with the corresponding S-box number. The higher-order
function \textit{foldl} is used with the function
\textit{serpentFold} to fold the input data block over the zipped
list of pairs. In other words, the function \textit{foldl}
replicates the required 31 rounds in a pipelined fashion. The
final round is carried out by XORing the output from the $31^{st}$
round with the $32^{nd}$ set of subkeys \textit{(sKeys!!32)}, at
this point the result is passed to the function \textit{s7}. The
final ciphered output is the result of XORing the output from the
function \textit{s7} with the last set of subkeys
\textit{(sKeys!!32)}. The suggested formal functional
specification is as follows:

 {\small \begin{verbatim}
 serpentEncryptSeg :: [[SubKey]] ->
             [DataBlock]->[DataBlock]

 serpentEncryptSeg sKeys input =
  zipWith fullexor (sKeys!!32) (s7 xorOut))
  where
    xorOut = zipWith fullexor (sKeys!!31)
                              roundsOut

    roundsOut = foldl serpentFold input
     (zip (take 31
      (concat (copy1 [0,1,2,3,4,5,6,7] 5)))
        (take 31 sKeys))
\end{verbatim}}

A \textit{Serpent} fold, specified as the function
\textit{serpentFold}, inputs a data block and a pair corresponding
to a list of four subkeys and the corresponding S-box number
employed in that fold. The subkeys are zipped with the input,
passed to the corresponding S-box, and finally linearly
transformed using the function \textit{lTransfrom}. The input
S-box number is used to choose one of the available S-boxes listed
in the list of functions \textit{s}. A possible formalisation is
as follows:

 {\small \begin{verbatim}
 serpentFold :: [DataBlock] ->
        (Int, [SubKey]) -> [[Bool]]

 serpentFold input (i,skey) =
    lTransform ((s!!i)
        (zipWith fullexor skey input))
    where
    s = [s0, s1, s2, s3,
         s4, s5, s6, s7]
\end{verbatim}}

The function \textit{lTransform} linearly transforms a list of 4 inputs into a list of 4 outputs.
the transformation uses the left circular shift function \textit{shift} and the left shift
function \textit{lshift} as follows:

 {\small \begin{verbatim}
 lTransform :: [[Bool]] -> [[Bool]]
 lTransform [x0, x1, x2,x3] =
    [y0, y1, y2, y3]
  where
    [y0i, y2i, y0, y1, y2, y3] =
    mapWith [(shift 13), (shift 3),
                (shift 5), (shift 1),
                (shift 22), (shift 7)]
            [x0, x2, y0ii, y1i, y2ii, y3i]

    [y1i, y3i, y0ii, y2ii] =
        zipWith fullexor
            (zipWith fullexor
                [x1, y2i, y0i, y2i]
                [y0i, (lshift 3 y0i),
                             y1, y3])
            [y2i, x3, y3, (lshift 7 y1)])

 lshift :: Int -> [Bool] -> [Bool]
 lshift n ls =
 (drop n ls) ++ (copy False n)
\end{verbatim}}

%%%%%%%%%%%%%%%%%%%%%%%%%%%%%%%%%%%%%%%%%%%%%%%%%%%%%%%%%%%%%%%%%%%%%%%%%%%%%%%%%

\subsection{Algorithms Refinement to CSP}
For the key scheduling part we suggest two designs. The first
design implements the scheduling in a data-parallel fashion. The
second design economises the implementation by carefully removing
replication from one of the main building blocks. For the
encryption part, we suggest three designs. The first design
presents a fully pipelined network of rounds. The second design
uses only one stage from the pipeline suggested in the first
design. In this case inputs and outputs are refined to streams.
The third design leaves a flexible choice for the level of
parallelism, allowing control over the number of pipelined stages.

%%%%%%%%%%%%%%%%%%%%%%%%%%%%%%%%%%%%%%%%%%%%%%%%%%%%%%%%%%%%%%%%%%%%%%%%%%%%%%%%%

\subsubsection{Key Scheduling} At this development stage, we refine
each function from the specification of the key scheduling part.
In the following section, the two suggested designs are presented
and explained.

\paragraph{First Design}

The types used in the specification of the function \textit{keySchedule} are refined to a 256-bit
Integer item for the private key, and a vector of vectors of items of size $(33 \times 4)$ for the
output subkeys:

 {\small
 $keySchedule :: Int256 \rightarrow \lfloor\lfloor Int32 \rfloor_{4}\rfloor_{33}$
 }

The refinement implements the function \textit{keySchedule} as a
process \textit{KEYSCHEDULE}. According to the specification, the
first event to occur is the segmentation of the input key into
eight segments using a predefined process \textit{SEGS}. These
eight segments are passed to the process \textit{GENERATEWS}.

{\small \noindent $KEYSCHEDULE = ((PRD(32) \rhd SEGS) \gg_{8} STORE_{v}(ws));(GENERATEWS(8, ws))
\gg_{132}$ $(VMAP_{4}(VMAPWITH([S_{0}, S_{1}, S_{2}, S_{3}, S_{4}, S_{5}, S_{6}, S_{7}]))\parallel
S_{3})$ }

\noindent where,

{\small

\noindent $S_{0} \sqsubseteq s0$; $S_{1} \sqsubseteq s1$; $S_{2} \sqsubseteq s2$; $S_{3}
\sqsubseteq s3$; $S_{4} \sqsubseteq s4$; $S_{5} \sqsubseteq s5$; $S_{6} \sqsubseteq s6$; $S_{7}
\sqsubseteq s7$;}
\\

\begin{figure}
	\begin{center}
		\includegraphics [scale = 0.6] %[height=2in,width=2in,angle=0]
		{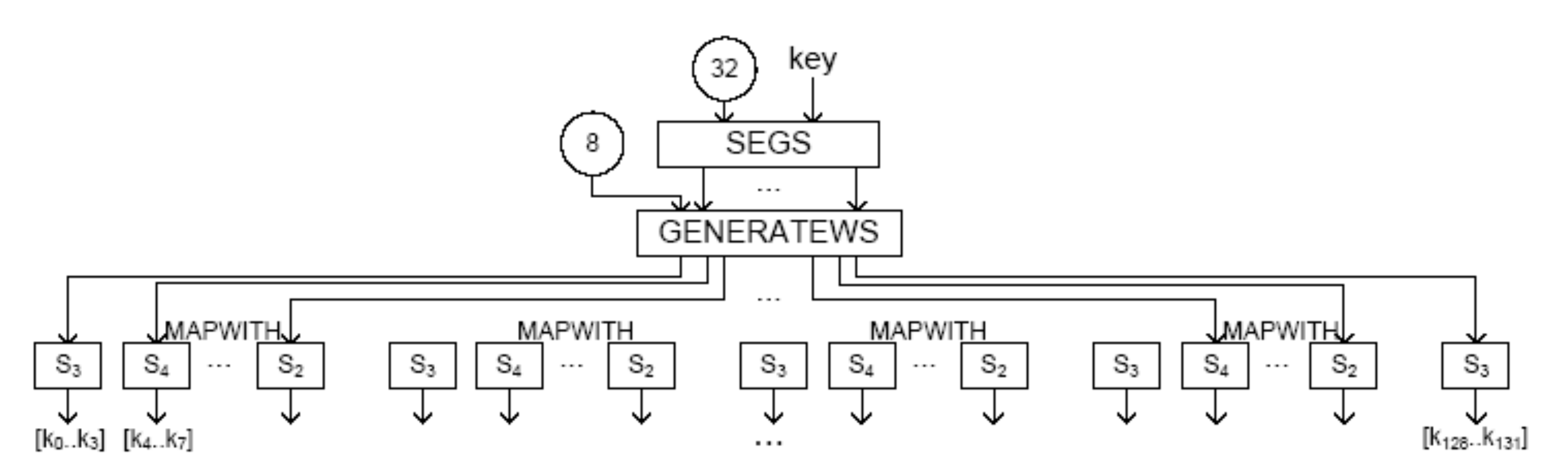}
		\caption{The process KEYSCHEDULE, first design}
		\label{KeySch1stD}
	\end{center}
\end{figure}

The higher-order process $(VMAP_{4})$ creates four parallel
instances of the process \textit{VMAPWITH}. In turn, 32 parallel
instances of the S-boxes processes is now available for parallel
computation. These 32 S-boxes process takes 128 items from the 132
generate prekeys in the process \textit{GENERATEWS}. The final
four prekeys are passed to a parallel instance of the process
\textit{S3}. The output from these parallel S-boxes processes is
the desired vector of 132-round subkeys. The process
\textit{KEYSCHEDULE} is depicted in Figure~\ref{KeySch1stD}.
\\

%Figure     \caption{The process KEYSCHEDULE, first design}

The function \textit{generateWs} could be refined as follows:

{\small
$generateWs :: Int32 \rightarrow \lfloor Int32 \rfloor_{8} \rightarrow \lfloor Int32
\rfloor_{132}$
\\

\noindent $generateWs \sqsubseteq GENERATEWS$

\noindent $GENERATEWS(i, ws)=$

$if (7 < i < 140)$

$then$ \ \ $WsD(i, ws) \gg StoreItem(wsd);$

\ \ \ \ \ \ $GENERATEWS(i+1, ws \concat [wsd])$

$else$ \ \ $PRD(ws)$}
\\

Unrolling the above recursive implementation for $GENERATEWS(8, ws)$:
\\

{\small

\noindent$GENERATEWS(8, ws)=$

\noindent$WsD(8, ws) \gg STORE_{v}(wsd); GENERATEWS(9, ws \concat [wsd]);$

\noindent$WsD(9, ws) \gg STORE_{v}(wsd); GENERATEWS(10, ws \concat [wsd]);$

\noindent$.$

\noindent$.$

\noindent$.$

\noindent$WsD(139, ws) \gg STORE_{v}(wsd); GENERATEWS(140, ws \concat [wsd]); PRD(ws)$}
\\

This could be done as:
\\

{\small

\noindent$GENERATEWS(8, ws)=$

\noindent$for(i = 8; i < 140; i++)\{$

\noindent$WsD(i, ws) \gg StoreItem(wsd);\}$}

\noindent where,

{\small

\noindent $WsD(i, ws) = out!(\ll_{11}(ws[i-8]\oplus ws[i-5]\oplus
ws[i-3]\oplus ws[i-1]\oplus (9e3779b9_{hex}) \oplus (i-8)))$}

%%%%%%%%%%%%%%%%%%%%%%%%%%%%%%%%%%%%%%%%%%%%%%%%%%%%%%%%%%%%%%%%%%%%%%%%%%%%%%%%%

\begin{figure}
	\begin{center}
		\includegraphics [scale = 0.65] %[height=2in,width=2in,angle=0]
		{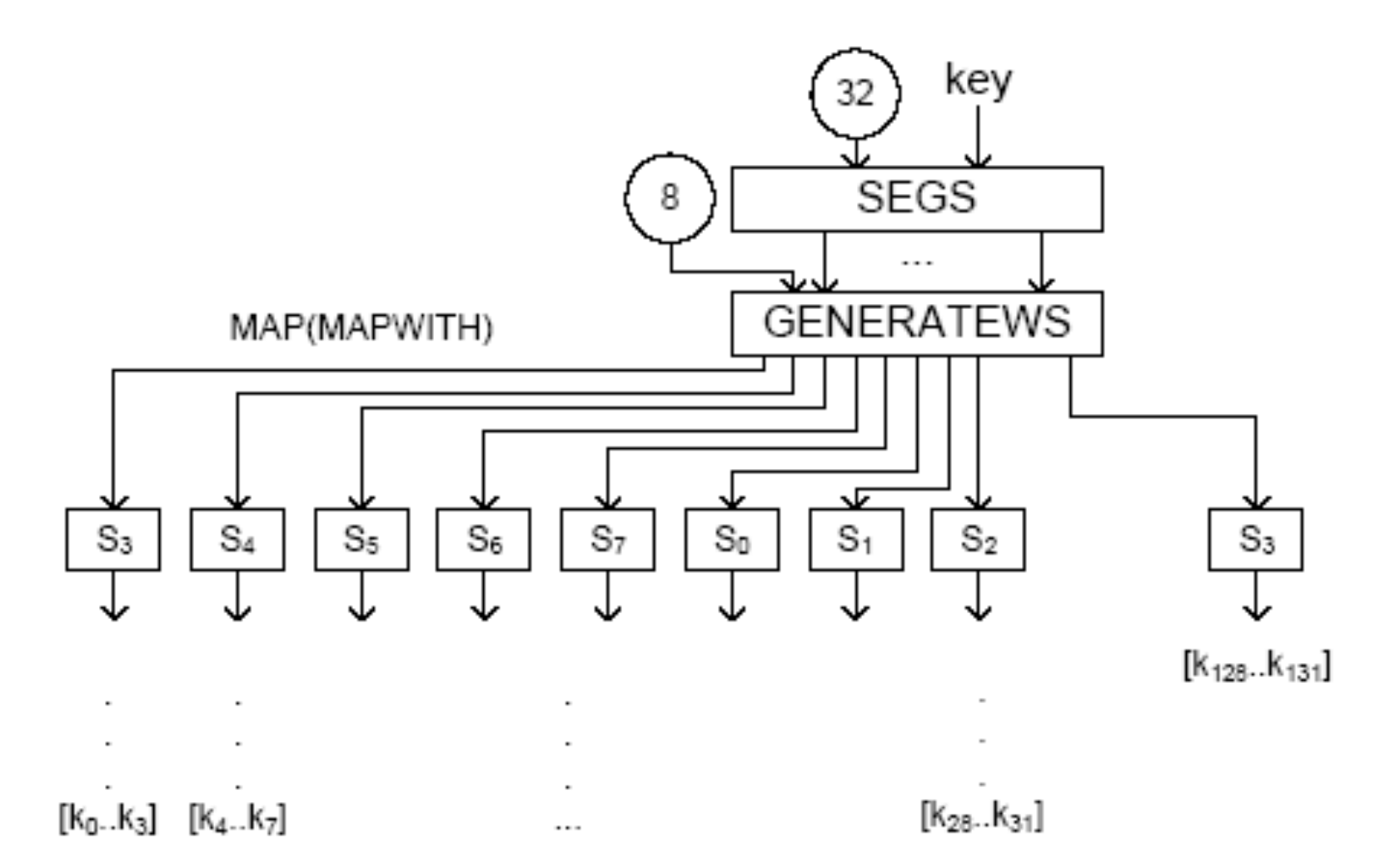}
		\caption{The process KEYSCHEDULE, second design with replication reduced}
		\label{KeySch2ndD}
	\end{center}
\end{figure}

\paragraph{Second Design} The second design intends to eliminate the replication in the S-boxes
computation processes. This leads to a smaller hardware circuit in the later stage as a trade for
the expected speed. The change from the first design is made by refining \textit{map} to its
stream setting. This implementation is depicted in Figure~\ref{KeySch2ndD} and described in the
following \textit{CSP} network:
\\

{\small
 $KEYSCHEDULE = ((32 \rhd SEGS) \gg_{8} STORE_{v}(ws));$

$(GENERATEWS(8, ws)) \gg$

$(SMAP(VMAPWITH([S_{0}, S_{1}, S_{2}, S_{3}, S_{4}, S_{5}, S_{6}, S_{7}]))\parallel S_{3})$}

%Figure     \caption{The process KEYSCHEDULE, second design with replication reduced}

%%%%%%%%%%%%%%%%%%%%%%%%%%%%%%%%%%%%%%%%%%%%%%%%%%%%%%%%%%%%%%%%%%%%%%%%%%%%%%%%%

\subsubsection{Serpent Block Cipher}
The current refinement is done in three different designs. The
process responsible for a single block ciphering is
\textit{SERPENTESEG}, the refinement of the function
\textit{serpentEncryptSeg}. The input data items, for instance,
could be passed as a stream of vectors of four 32-bit data items
to the encrypting block \textit{SERPENTESEG}. The output is
refined also to a stream of items as follows:

{\small

$serpentEncrypt(key) :: \langle \lfloor Int32 \rfloor_{4} \rangle \rightarrow \langle \lfloor
Int32 \rfloor_{4} \rangle$}

Consequently, we suggest the following refinement employing the higher-order process
\textit{SMAP}. The \textit{key}, in this case, is passed as an argument to the process
\textit{SERPENTENCRYPT}.

{\small

$SERPENTENCRYPT(key) = KEYSCHEDULE(Key) \gg SMAP(SERPENTESEG)$}

A multi-way \textit{Serpent} encryption version is implemented as follows:
\\

{\small
$serpentEncrypt(key) :: \langle \lfloor \langle \lfloor Int32 \rfloor_{4} \rangle
\rfloor_{n} \rangle \rightarrow \langle \lfloor \langle \lfloor Int32 \rfloor_{4} \rangle
\rfloor_{n} \rangle$
\\

$SERPENTENCRYPT(key) = KEYSCHEDULE(key) \gg$ \\
$SMAP(VMAP_{n}(SMAP(SERPENTESEG)))$}\\

where the value of \textit{n} is limited by the ability to realise this network on the available
hardware in the following stage. The following three designs are suggested for the implementation
of the process \textit{SERPENTESEG}.

%%%%%%%%%%%%%%%%%%%%%%%%%%%%%%%%%%%%%%%%%%%%%%%%%%%%%%%%%%%%%%%%%%%%%%%%%%%%%%%%%

\begin{figure}
	\begin{center}
		\includegraphics [scale = 0.55] %[height=2in,width=2in,angle=0]
		{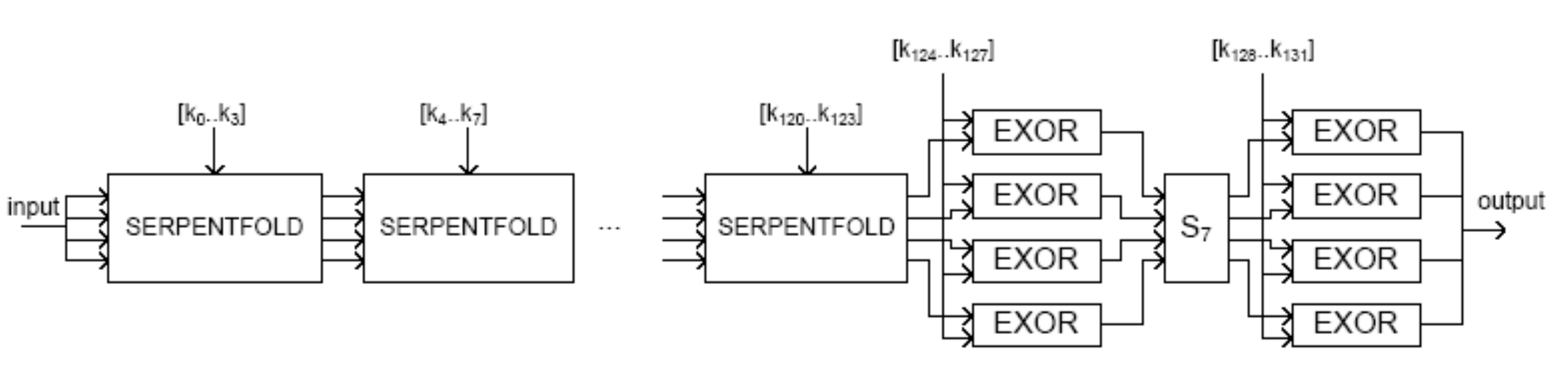}
		\caption{The process SERPENTESEG, first pipelined design}
		\label{Serpent1stD}
	\end{center}
\end{figure}

\paragraph{First Design} This design suggests a fully pipelined implementation of the \textit{Serpent}
encryption specification. The pipeline is constructed by
replicating the single round specified as the function
\textit{serpentFold}. The replication is done using the vector
setting refinement of the higher-order function \textit{foldl},
where the input is a vector of items. The input 132 subkeys are
distributed to the pipelined folds as shown in
Figure~\ref{Serpent1stD}. Also, the number of the round in use is
distributed to the pipelined folds. The output from the pipeline
is the input to the higher-order process \textit{VZIPWITH(EXOR)},
zipping it with a set of four subkeys. The result of zipping is
passed to an $S_{7}$ S-box process, whose output vector is zipped
again using another \textit{VZIPWITH(EXOR)} with the last
generated set of four subkeys. The \textit{CSP} description is as
follows:
\\

{\small

\noindent $serpentEncryptSeg = \lfloor Int32 \rfloor_{132} \rightarrow \langle \lfloor Int32
\rfloor_{4}\rangle \rightarrow \langle \lfloor Int32 \rfloor_{4}\rangle$
\\

\noindent $serpentEncryptSeg \sqsubseteq SERPENTESEG$

\noindent $SERPENTESEG = (BROADCAST_{3}([0..7]) \parallel (PRD([0..6]))) \rhd
(VVFOLDL(SERPENTFOLD) \parallel VZIPWITH_{4}(EXOR) \gg_{4} S_{7} \parallel VZIPWITH_{4}(EXOR))$}

\noindent where,

{\small $serpentFold \sqsubseteq SERPENTFOLD$}

%Figure    \caption{The process SERPENTESEG, first pipelined design}

The serpent fold is implemented as in the following:

{\small

\noindent $serpentFold ::  \lfloor Int32 \rfloor_{4} \rightarrow (Int3,  \lfloor Int32
\rfloor_{4}) \rightarrow  \lfloor Int32 \rfloor_{4}$}

{\small \noindent $SERPENTFOLD = (in?i \rightarrow SKIP); VZIPWITH_{4}(EXOR) \gg S_{i} \gg
LTRANSFORM$}

\noindent where

{\small $lTransform \sqsubseteq LTRANSFORM$}

The linear transformation function \textit{lTransform} is refined to the process
\textit{LTRANSFORM}. The input and output are refined as a vector of items as follows:

{\small $lTransform :: \lfloor Int32 \rfloor_{4} \rightarrow \lfloor Int32 \rfloor_{4}$}

The process \textit{LTRANSFORM} is implemented as follows:
\\

{\small

$LTRANSFORM = (\interleave_{i=0}^{i=3}in[i]? x[i] \rightarrow SKIP);$

$LSHIFT(3) \parallel LSHIFT(7) \parallel$

$(VZIPWITH_{4}(EXOR) \gg_{4} VZIPWITH_{4}(EXOR)) \parallel$

$VMAPWITH([SHIFT(1), SHIFT(13), SHIFT(3),$

$SHIFT(5), SHIFT(1), SHIFT(22), SHIFT(7)])$}

%%%%%%%%%%%%%%%%%%%%%%%%%%%%%%%%%%%%%%%%%%%%%%%%%%%%%%%%%%%%%%%%%%%%%%%%%%%%%%%%%

\begin{figure}
	\begin{center}
		\includegraphics [scale = 0.5] %[height=2in,width=2in,angle=0]
		{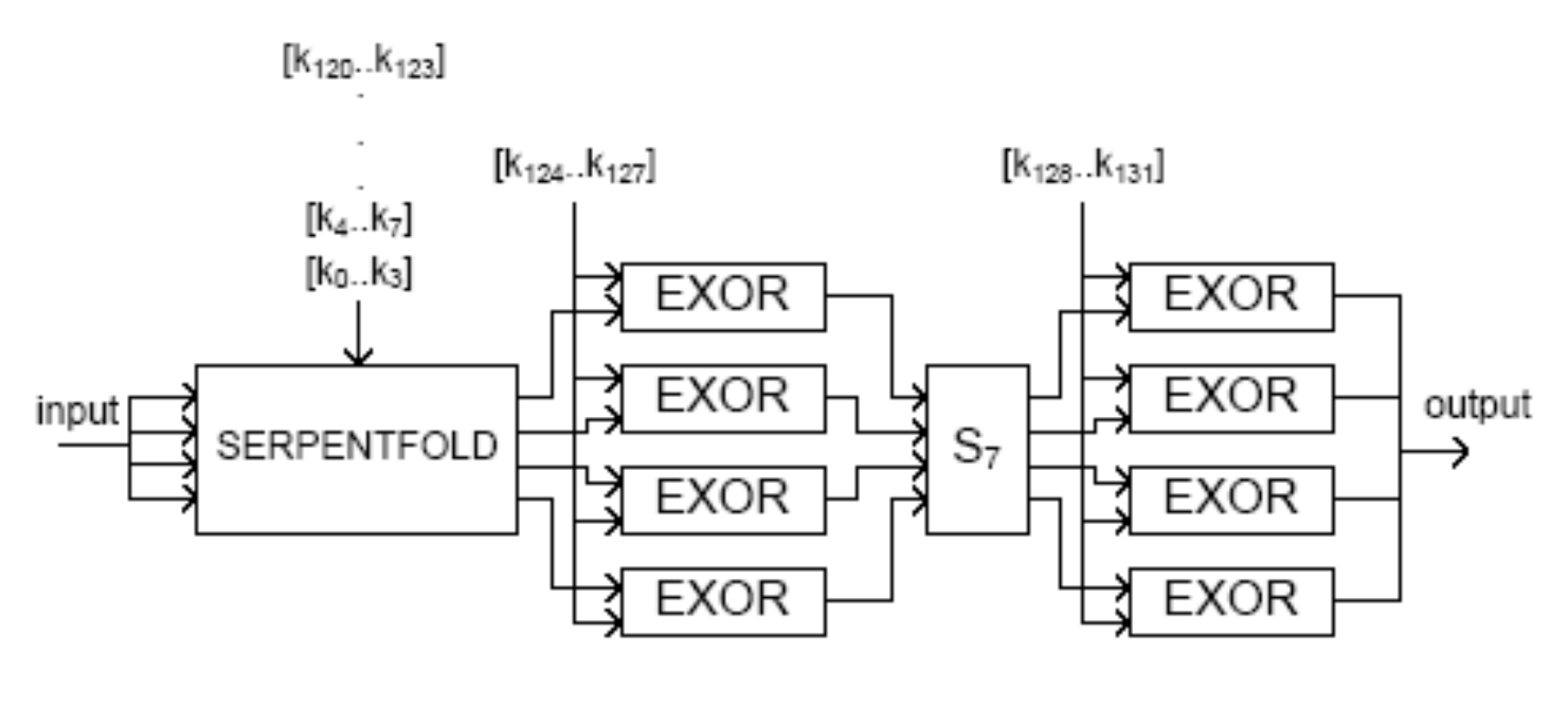}
		\caption{The process SERPENTESEG, second design with stream of subkeys}
		\label{Serpent2ndD}
	\end{center}
\end{figure}

\paragraph{Second Design}
In this design, the network component processes are still the
same, as shown in the first design, with a modification to the way
they communicate. The stream communication with the main process
\textit{SERPENTFOLD}, allows the elimination of copies of this
process using \textit{SVFOLDL} the stream refinement of
\textit{foldl}, where the input is a vector of items. The subkeys
distribution, at this point, are passed sequentially to the
process \textit{SERPENTFOLD}. Only the last two sets of subkeys
are produced as vectors to be used in the two similar parallel
processes $VZIPWITH_{4}(EXOR)$. This network is shown in
Figure~\ref{Serpent2ndD}. The \textit{CSP} description is as
follows:
\\

{\small
 $serpentEncryptSeg = \langle Int32 \rangle \rightarrow \langle \lfloor Int32
\rfloor_{4}\rangle \rightarrow \langle \lfloor Int32 \rfloor_{4}\rangle$
\\

$SERPENTESEG = (BROADCAST_{3}([0..7]) \parallel (PRD([0..6]))) \rhd (SVFOLDL(SERPENTFOLD)
\parallel VZIPWITH_{4}(EXOR) \gg_{4} S_{7} \parallel VZIPWITH_{4}(EXOR))$}
\\

%Figure     \caption{The process SERPENTESEG, second design with stream of subkeys}

%%%%%%%%%%%%%%%%%%%%%%%%%%%%%%%%%%%%%%%%%%%%%%%%%%%%%%%%%%%%%%%%%%%%%%%%%%%%%%%%%

\begin{figure}
	\begin{center}
		\includegraphics [scale = 0.5] %[height=2in,width=2in,angle=0]
		{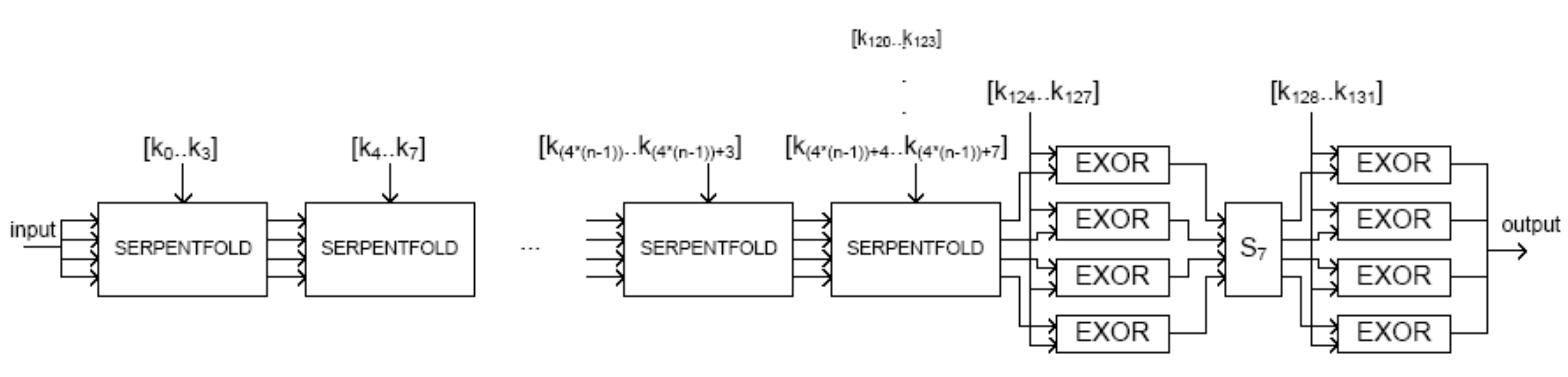}
		\caption{The process SERPENTESEG, third partially pipelined design}
		\label{Serpent3rdD}
	\end{center}
\end{figure}

\paragraph{Third Design}
Based on the above suggested implementations, this design composes both a
pipelined part and a stream-based part to build the final desired \textit{Serpent} network. This
implementation is shown in Figure~\ref{Serpent3rdD} and done as follows:
\\

{\small \noindent $serpentEncryptSeg = \langle Int32 \rangle \rightarrow \lfloor Int32
\rfloor_{132} \rightarrow \langle \lfloor Int32 \rfloor_{4}\rangle \rightarrow \langle \lfloor
Int32 \rfloor_{4}\rangle$
\\

\noindent$SERPENTESEG = (BROADCAST_{3}([0..7]) \parallel (PRD([0..6])))$

\noindent$\rhd (((PRD(n) \rhd VVFOLDL(SERPENTFOLD)) \parallel$

\noindent$SVFOLDL(SERPENTFOLD)) \parallel VZIPWITH_{4}(EXOR) \gg_{4}$

\noindent$S_{7} \parallel VZIPWITH_{4}(EXOR))$}

%% Figure     \caption{The process SERPENTESEG, third partially pipelined design}

%%%%%%%%%%%%%%%%%%%%%%%%%%%%%%%%%%%%%%%%%%%%%%%%%%%%%%%%%%%%%%%%%%%%%%%%%%%%%%%%%

\subsection{Reconfigurable Hardware Implementations}
The part of the hardware implementation included in this section is aimed to show samples of the
implemented code. We put some emphasis on some code segments, where we could not base the
implementation from the previous stage in a straightforward manner. Remember that the main reason
behind the faced coding difficulties resides in the level of generality of the constructs to be
implemented.

The following macros are for the two designs of key scheduling.
The first macro \textit{KeySchfedule1st} outputs the subkeys as a
vector of vectors of vectors of items from the macros
\textit{GenerateWsVOVOV} and \textit{S3}.

 {\small \begin{verbatim}

 macro proc KeySchedule1st
 (keyIn, KssOutVOVOV, lastksV){
 .
 .
 .
 par{
  Segs(keyIn, segmentsOut);

  GenerateWsVOVOV
  (segmentsOut, WsOutVOVOV, lastwsV);

  VMap
  (WsOutVOVOV, 4, KssOutVOVOV, VMapWithSs);
  S3(lastwsV, lastksV);}
\end{verbatim}}

The macro for the second design with its stream implementation is
as follows:

 {\small \begin{verbatim}
 macro proc KeySchedule2nd
    (keyIn, KssOutSOVOV, lastksV) {
 .
 .
 .
 par{
  Segs(keyIn, segmentsOut);

  GenerateWsSOVOV
    (segmentsOut, WsOutSOVOV, lastwsV);

  Map(WsOutSOVOV, KssOutSOVOV, VMapWithSs);
  S3(lastwsV, lastksV);}}
\end{verbatim}}

The stream-version macro implementing the process
\textit{GenerateWs} is shown in the following code section.  In
this macro a $140$ (32-bit Integer) elements array \textit{ws} is
used to store the generated prekeys. This means occupying a large
area from the targeted \textit{FPGA}. An alternative
implementation is to use the available internal \textit{RAM}, so,
this would dramatically save the needed space. The \textit{RAM}
property of allowing only one access to it at once (read or write
at a time) imposes some restrictions. For instance, the production
of the final calculated prekeys should be done as stream of items
instead of a stream of vectors of vectors of items. Both cases are
shown in the following code sections:

 {\small \begin{verbatim}
macro proc GenerateWsSOVOV
(wsIn, wsOutSOVOV, lastwss) {
.
.
.
Int32 ws[140];

par(j = 0; j < 8; j++){
 jTemp[j] = 0@j;
 wsIn.elements[j].channel ?
        ws[jTemp[j]];}

PHI = 0x9e3779b9;

for(i = 8; i < 140; i++){
    iTemp = 0@i;
    wTemp = ws[i-3]^ws[i-5]^
            ws[i-8]^ws[i-1]^
            PHI^(iTemp-8);

    par{
        ProduceItem(wItem, wTemp);
        Shift(wItem, 11, sOut);
        StoreItem(sOut, ws[i]);}
    if (i == 139){
    break;}}

ProduceSOVOVOItemsFromArrayWithOffset
    (wssOutSOVOV, 4, 8, 4, ws, 8);
par{
 lastwss.elements[0].channel ! ws[136];
 lastwss.elements[1].channel ! ws[137];
 lastwss.elements[2].channel ! ws[138];
 lastwss.elements[3].channel ! ws[139];}}
\end{verbatim}}

The second version is as follows:

 {\small \begin{verbatim}
 macro proc GenerateWsRam
    (wsIn, wssOut) {
 .
 .
 .
 ram Int32 ws[140];

 par(j = 0; j < 8; j++){
 jTemp[j] = 0@j;
 wsIn.elements[j].channel ?
    ws[jTemp[j]];}

 PHI = 0x9e3779b9;

 for(i = 8; i < 140; i++){
    iTemp = 0@i;
    wTemp = ws[i-1]^PHI^(iTemp-8);
    wTemp1 = wTemp ^ ws[i-8];
    wTemp2= wTemp1^ws[i-5];
    wTemp3 = wTemp2^ws[i-3];
    par{
        ProduceItem(wItem, wTemp3);
        Shift(wItem, 11, sOut);
        StoreItem(sOut, ws[i]);}
    if (i == 139){
    break;}}
 ProduceStreamOfItems(wsOut, 140, ws);}
\end{verbatim}}

The use of an \textit{FPGA's} on-chip memory is constrained with
its supported memory capabilities and corresponding
\textit{Handel-C} compilation options. The available sophisticated
\textit{SelectRAM} memory hierarchy available on the used
\textit{Virtix-E FPGA} supports \textit{True Dual-Port BlockRAMs}
and Distributed \textit{RAMs}. However, \textit{Handel-C}
declaration of an array is equivalent to declaring a number of
variables. Each entry in an array may be used exactly like an
individual variable, with as many reads, and as many writes to a
different element in the array as required within a clock cycle.
Arrays are more efficient to implement in terms of concurrent
access required by fast pleasantly parallel designs. Arrays are
implemented using the available logic blocks in an \textit{FPGA}
(Slices in the case of \textit{Xilinx} devices). \textit{RAMs},
are normally more efficient to implement in terms of hardware
resources than arrays since they use the on-chip \textit{RAM}
blocks. \textit{RAMs}, would allow one location to be accessed in
any one clock cycle.

To take the advantage of an available multi-port memory blocks,
one can use the \textit{mpram} declaration in \textit{Handel-C}
instead of \textit{ram}. A design that uses an \textit{mpram} with
two ports would outperform the sequential design in terms of
speed, but still replications of some processes would be necessary
to cope with the doubled amount of information retrieved. A design
that uses a dual-ported memory to store a list should have refined
the list as a stream of vectors of two elements in the description
stage.

Before we present parts of the realization of the encryption
designs, we note the solution we suggest for implementing the
higher-order process \textit{VMAPWITH} with a list of different
processes. The macro \textit{VMapWith} needs to map a list of
macros to a list of items. The problem we faced is for how to pass
a list of macros as an argument to the macro \textit{VMapWith}. A
best case scenario is having the following code implementation:

 {\small \begin{verbatim}

 macro proc VMapWith
    (vIn, , vProcesses, vOut, n){

 par(i = 0, i < n, i++){
   vProcesses[i]
    (vIn.elements[i],
       vOut.elements[i]);}}
\end{verbatim}}

The vector of macros \textit{vProcesses} passing to the macro \textit{VMapWith} is not supported
in the current version of \textit{Handel-C}. A second possible form for a possible implementation
in \textit{Handel-C} is as follows:

 {\small \begin{verbatim}

 macro proc VMapWith
 (vIn, P1, P2,..., Pn, vOut, n){

 par{
  P1(vIn.elements[i], vOut.elements[i]);
  P2(vIn.elements[i], vOut.elements[i]);
  .
  .
  .
  Pn(vIn.elements[i], vOut.elements[i]);}}
\end{verbatim}}

A step forward in the code generation leads to the third possible
form of implementation. This form would fit the calling of the
process \textit{VMAPWITH} from another higher-order macro as had
been done in:

{\small \begin{verbatim}
 VMap(WsOutVOVOV, 4,
      KssOutVOVOV, VMapWithSs);
\end{verbatim}}

This suggests the removing of the zipped-with macro names from the arguments lists in the macro
procedure definition as follows:

 {\small \begin{verbatim}

 macro proc VMapWithPs(vIn, vOut, n){

 par{
  P1(vIn.elements[i], vOut.elements[i]);
  P2(vIn.elements[i], vOut.elements[i]);
  .
  .
  .
  Pn(vIn.elements[i], vOut.elements[i]);}}
\end{verbatim}}

A possible solution to such a limitation is, again, the
availability of a preprocessor automatically generating the
allowed implementation from the best case scenario presented. For
the case of mapping with the list of S-boxes macros; the code is
as follows:

 {\small \begin{verbatim}
 macro proc VMapWithSs(vIn, vOut){
 par{
  S3(vIn.elements[0], vOut.elements[0]);
  S2(vIn.elements[1], vOut.elements[1]);
  S1(vIn.elements[2], vOut.elements[2]);
  S0(vIn.elements[3], vOut.elements[3]);
  S7(vIn.elements[4], vOut.elements[4]);
  S6(vIn.elements[5], vOut.elements[5]);
  S5(vIn.elements[6], vOut.elements[6]);
  S4(vIn.elements[7], vOut.elements[7]);}}
\end{verbatim}}

For the encryption part, we include the implementation done for the third design. Whereby, a
combination of parallel and sequential fold are employed with vector of items as input. Based on
the \textit{CSP} implementation, the macro \textit{EncryptSegsVVandSV} is implemented as follows:

{\small \begin{verbatim}
 macro proc EncryptSegsVVandSV
     (input, sKeysVOV, VRnds,
    sKeysSOV, finalKeys, output) {

 par{
  VVFoldL(sKeysVOV, output1, 4,
    NParalRnds, SerpentFold, input);

  SVFoldL(sKeysSOV, 4, output2,
    SerpentFold, output1, 4, NParalRnds);

  VZipWith(4, output2,
    finalKeys.elements[0], output3, EXOR);

  S7(output3, output4);

  VZipWith(4, output4,
    finalKeys.elements[1], output, EXOR);}}
\end{verbatim}}

The macro \textit{SerpentFold} implements its corresponding process as follows:

 {\small \begin{verbatim}

 macro proc SerpentFold
     (input, i, sKeys, output) {

 VectorOfItems (vOut, 4, Int32);
 VectorOfItems (output1, 4, Int32);

 par{
    par{
    VZipWith(input, sKeys, vOut, EXOR);}

    if(i==0)
        S0(vOut, output1);
    else if(i==1)
        S1(vOut, output1);
    else if(i==2)
        S2(vOut, output1);
    else if(i==3)
        S3(vOut, output1);
    else if(i==4)
        S4(vOut, output1);
    else if(i==5)
        S5(vOut, output1);
    else if(i==6)
        S6(vOut, output1);
    else if(i==7)
        S7(vOut, output1);

    LinearTransformation
        (output1, output);}}
\end{verbatim}}

%%%%%%%%%%%%%%%%%%%%%%%%%%%%%%%%%%%%%%%%%%%%%%%%%%%%%%%%%%%%%%%%%%%%%%%%%%%%%%%%%

\section{General Evaluation} \label{GE}

\begin{comment}
In this paper, we have demonstrated a methodology that can produce
intuitive, high-level specifications of algorithms in the
functional programming style. The development continues by
deriving efficient, parallel implementations described in
\textit{CSP} and realized using \textit{Handel-C} that can be
compiled into hardware on an \textit{FPGA}. We have provided a
concrete study that exploited both data and pipelined parallelism
and the combination of both. The implementation was achieved by
combining behavioral implementations 'off-the-shelf' of commonly
used components that refine the higher-order-functions which form
the building blocks of the starting functional specification.
\end{comment}

In this paper, the contribution of the presented work could be
found in many aspects. Some additions were crucial to the
realization step of the method so that it can cope with real-life
complex areas of applications. A famous algorithm from
cryptography has been targeted as a test case that has given a
clear idea about the practical use of the methodology. Reusable
libraries are created at all levels of development. The
availability of such libraries supports and facilitates the
development in general. The created libraries for the different
studies from cryptography are highly reusable for developing other
cryptographic algorithm. This might include the introduction of
new components to the libraries, or slightly modifying the
available ones. According to these points, we stress the following
aspects:

The development is originated from a specification stage, whose
main key feature is its powerful \textbf{higher-level of
abstraction}. During the specification, the isolation from
parallel hardware implementation issues allowed for deep
concentration on the specification details. Whereby, for the most
part, the style of specification comes out in favor of using
higher-order functions. Two other inherent advantages for using
the functional paradigm are \textbf{clarity} and
\textbf{conciseness} of the specification. This was reflected
throughout all the presented studies. At this level of
development, the \textbf{correctness} of the specification is
insured by construction from the used correct building blocks. The
implementation of the formalized specification is tested under
\textit{Haskell} by performing random tests for every level of the
specification.

The correctness will be carried forward to the next stage of
development by applying the provably correct rules of refinement.
The available pool of refinement formal rules enables a high
degree of \textbf{flexibility} in creating parallel designs. This
includes the capacity to divide a problem into completely
independent parts that can be executed simultaneously (pleasantly
parallel). Conversely, in a nearly pleasantly parallel manner, the
computations might require results to be distributed, collected
and combined in some way. Remember at this point, that the
refinement steps are done by combining off-the-shelf
\textbf{reusable} instances of basic building blocks.

\section{Performance Analysis} \label{PA}

In this section we show the testing results of mapping the
designs, analyzing their timing, and showing the speeds as
measured for testing the \textit{RC-1000} board from the used
\textit{P4} machine. The fully-pipelined design was over-mapped,
thus, the following presented speeds are for the remaining
designs. Note that in the suggested \textit{Serpent}
implementations, the finest grains of basic building blocks are
refined as processes rather than using \textit{Handel-C}
operators. Thus, an increase in communications cost between
processes is found.

In Table~\ref{TblEnSKeys}, we show the testing results of the encryption subkeys generation. The
keys generation (second design) runs with a throughput of $96$ Mbps occupying $13097$ Slices, i.e.
$68\%$ of the \textit{FPGA} area.

\begin{table} 
	\caption{Testing results of \textit{Serpent} encryption subkeys generation}
	\label{TblEnSKeys}
	\begin{center}
		\includegraphics [scale=0.5]%[height=6in,width=5in,angle=0]
		{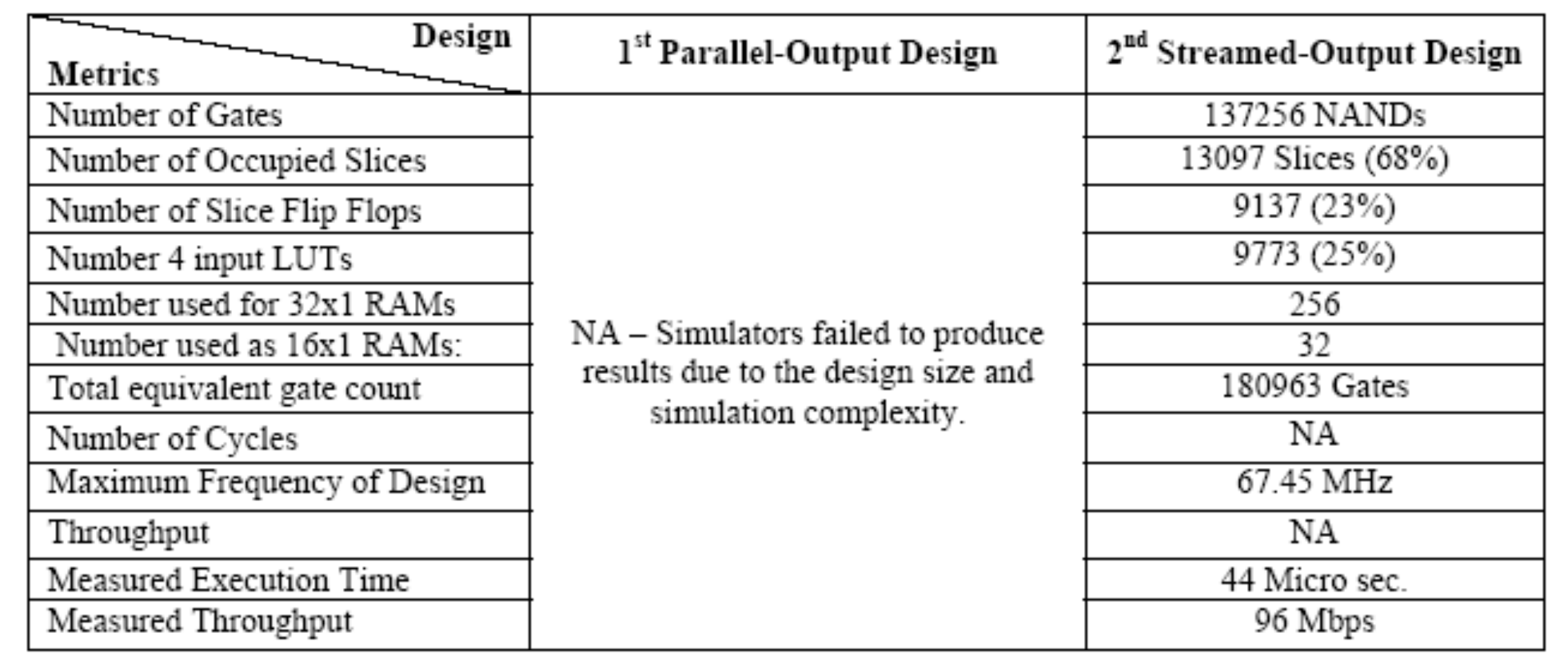}
	\end{center}
\end{table}

%table     \caption{Testing results of \textit{Serpent} encryption subkeys generation}

As shown in Table~\ref{TblEnc}, the testing results of the \textit{Serpent} second and third
designs are included, while the first design failed to compile with its large gates count. The
maximum achieved parallelism was in running the third design with 2 parallel folds and a third
performing the remaining 29 sequential folds. This implementation has a throughput of $12.21$ Mbps
occupying an area of 19198 Slices ($99\%$ of the available \textit{FPGA} area). The second design
with its sequential single fold implementation achieved throughput of $12.15$ Mbps with an area of
$12291$ Slices.

%Table     \caption{Testing results of \textit{Serpent} encryption}

\begin{table}
	\caption{Testing results of \textit{Serpent} encryption}
	\label{TblEnc}
	\begin{center}
		\includegraphics [scale=0.6]%[height=6in,width=5in,angle=0]
		{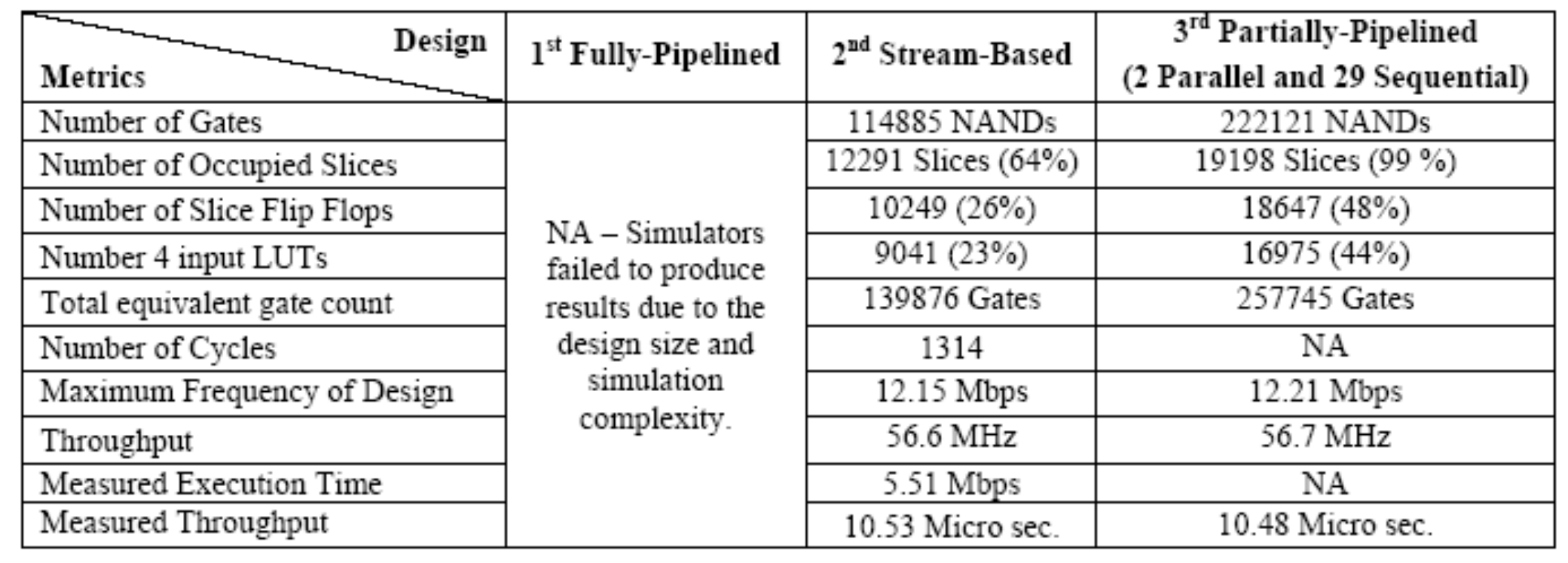}
	\end{center}
\end{table}

In Table~\ref{TblComparisons}, we include some results from
literature mapping the same algorithms onto \textit{FPGAs}. The
high-speeds achieved for the suggested optimised implementations
is very clear, as compared to our high-level (un-optimised)
implementation (yet) - from performance perspective. The shown
results include a high-speed implementations for the
\textit{Serpent} ($333$ Mbps) presented by Elbirt et al \cite{s2}.
Gaj et al in \cite{Rj6} presented another high-speed
implementation for the \textit{Serpent} ($431.4$ Mbps).

\begin{table}
	\caption{Comparisons among similar \textit{FPGA} systems implementing optimized \textit{Serpent}}
	\label{TblComparisons}
	\begin{center}
		\includegraphics [scale=0.8] %[height=5in,width=4in,angle=0]
		{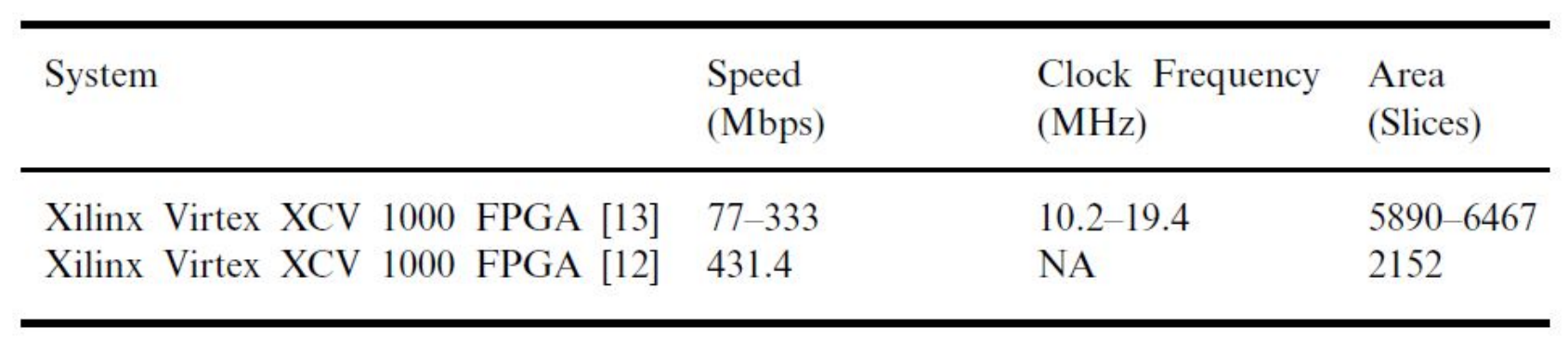}
	\end{center}
\end{table}

In Table~\ref{TblComparisonsNew} \cite{s12,s13,s14} we compare the
number of cycles for different hardware implementations of the
\textit{Serpent} including a number of microprocessor-based
implementations.

%table    \caption{Comparisons among similar \textit{FPGA} systems implementing optimized \textit{Serpent}}
%table    \caption{Comparisons among different hardware systems, with respect to number of clock cycles, implementing the \textit{Serpent}}

\begin{table}
	\caption{Comparisons among different hardware systems, with respect to number of clock cycles, implementing the \textit{Serpent}}
	\label{TblComparisonsNew}
	\begin{center}
		\includegraphics [scale=0.65] %[height=5in,width=4in,angle=0]
		{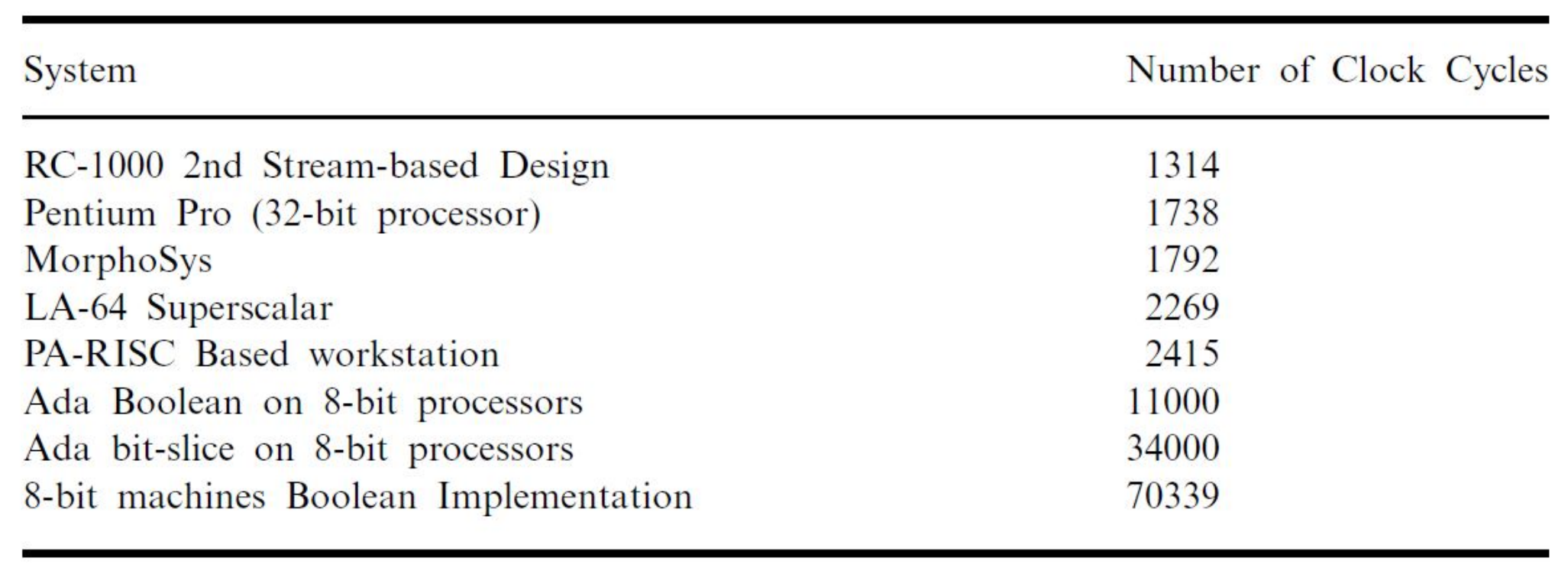}
	\end{center}
\end{table}

The higher-level development caused high replication in using
basic building blocks, and more clearly their communications. Many
instances of \textit{PRODUCE} and \textit{STORE} processes caused
the high use of intermediate variables. Other processes were used
for structuring data in the format corresponding to their
functional definitions. For instance, to collect some vectors of
subkeys and produce them as a vector of vectors of vectors of
items. Such use also plays a big role in occupying larger silicon
area after realization.

If we consider the implementation of an algorithm without using
our proposed method, we might implement the whole design with a
small number of macros and  minimum use of communications.
Moreover, possible handmade enhancements could be done with the
aid of shared variables. This would undoubtedly reduce the cost
paid for communicating parallel processes implementation and might
lead to a more economical realization and less congested design
with a higher frequency. This certainly comes as \textit{quid pro
quo} for the step-wise development.

%%%%%%%%%%%%%%%%%%%%%%%%%%%%%%%%%%%%%%%%%%%%%%%%%%%%%%%%%%%%%%%%%%%%%%%%%%%%%%%%%%%%%%%%%%%%%%%%%%

\section{Conclusion} \label{Con}

Mapping parallel versions of algorithms onto hardware could
enormously improve computational efficiency. Recent advances in
the area of reconfigurable computing came in the form of
\textit{FPGAs} and their high-level \textit{HDLs} such as
\textit{Handel-C}. In this paper, we build on these recent
technological advances by presenting, demonstrating and examining
a high-level hardware development method. The used method creates
a functional specification of an algorithm without defining
parallelism. Correspondingly, an efficient parallel implementation
is derived in the form of \textit{CSP} network of processes.
Accordingly, we create diffident parallel implementations in
\textit{Handel-C}. The presented work included theory and
practices about the suggested methodology. In this paper, we
observed a case study from applied cryptography, namely the
\textit{Serpent} algorithm. The encryption block ciphers and key
expansions were addressed. The correctness, conciseness and
clarity of the specification is emphasized. The systematic and
flexible refinements of the specification allowed the reasoning
about various implementations with different degrees of
parallelism for each case. The described designs ranged from
fully-pipelined, partially-pipelined, to streamed input and output
implementations. At this stage, the realization using
\textit{Handel-C} is presented, emphasizing some code segments
which tackled different noted implementation pitfalls. Future work
includes extending the theoretical pool of rules for refinement,
the investigation of automating the development processes, and the
optimization of the realization for more economical
implementations with higher throughput.

%%%%%%%%%%%%%%%%%%%%%%%%%%%%%%%%%%%%%%%%%%%%%%%%%%%%%%%%%%%%%%%%%%%%%%%%%%%%%%%%%%%%%%%%%%%%%%%%%%%%%%%%%%%%%%%%%

\section*{Acknowledgment}

% optional entry into table of contents (if used)
%\addcontentsline{toc}{section}{Acknowledgment}

I would like to thank Dr. Ali Abdallah, Prof. Mark Josephs, Prof. Wayne Luk, Dr. Sylvia Jennings,
and Dr. John Hawkins for their insightful comments on the research which is partly presented in
this paper.

%%%%%%%%%%%%%%%%%%%%%%%%%%%%%%%%%%%%%%%%%%%%%%%%%%%%%%%%%%%%%%%%%%%%%%%%%%%%%%%%%%%%%%%%%%%%%%%%%%%%%%%%%%%%%%%%%



% http://www.ctan.org/tex-archive/macros/latex/contrib/supported/IEEEtran/

%download BibTeX files

%\bibliographystyle{IEEEtranS}

\bibliographystyle{elsart-num}


\bibliography{serpent}

\end{document}